\documentclass[12pt]{article}

\pdfoutput=1

\usepackage{amssymb}
\usepackage{amsmath}

\usepackage{graphicx}
\usepackage{amsfonts}
\usepackage{hyperref}

\def\unit{\relax{\rm 1\kern-.26em I}}
\def\nada{\relax{\rm 0\kern-.30em l}}

\makeatletter
\renewcommand\section{\@startsection {section}{1}{\z@}%
                                 {-3.5ex \@plus -1ex \@minus -.2ex}
                                   {2.3ex \@plus.2ex}%
                                   {\normalfont\large\bfseries}}
\renewcommand\subsection{\@startsection{subsection}{2}{\z@}%
                                   {-3.25ex\@plus -1ex \@minus -.2ex}%
                                     {1.5ex \@plus .2ex}%
                                     {\normalfont\bfseries}}
\renewcommand\subsubsection{\@startsection{subsubsection}{3}{\z@}%
                                   {-3.25ex\@plus -1ex \@minus -.2ex}%
                                     {1.5ex \@plus .2ex}%
                                     {\normalfont\itshape}}
\makeatother

\newcommand{\be}{\begin{equation}}
\newcommand{\ee}{\end{equation}}
\newcommand{\bea}{\begin{eqnarray}}
\newcommand{\eea}{\end{eqnarray}}
\newcommand{\barr}{\begin{array}}
\newcommand{\earr}{\end{array}}

\def\beq{\begin{equation}}
\def\eeq{\end{equation}}
\def\be{\begin{equation}}
\def\ee{\end{equation}}
\def\bea{\begin{eqnarray}}
\def\eea{\end{eqnarray}}

\DeclareRobustCommand{\SkipTocEntry}[4]{}

\begin{document}

\begin{titlepage}

\setcounter{page}{1} \baselineskip=19.5pt \thispagestyle{empty}

\begin{flushright}
DESY-12-106\\
\end{flushright}
\vfil

\begin{center}
{\LARGE Tensor modes on the string theory landscape}

\end{center}
\bigskip\

\begin{center}
{\large Alexander Westphal}
\end{center}

\begin{center}
\textit{Deutsches Elektronen-Synchrotron DESY, Theory Group, D-22603 Hamburg, Germany}
\end{center} \vfil

\noindent  We attempt an estimate for the distribution of the tensor mode fraction $r$ over the landscape of vacua in string theory. The dynamics of eternal inflation and quantum tunneling lead to a kind of democracy on the landscape, providing no bias towards large-field or small-field inflation regardless of the class of measure. The tensor mode fraction then follows the number frequency distributions of inflationary mechanisms of string theory  over the landscape. We show that an estimate of the relative number frequencies for small-field vs large-field inflation, while unattainable on the whole landscape, may be within reach as a regional answer for warped Calabi-Yau flux compactifications of type IIB string theory.

\vfil
\begin{flushleft}
June 19, 2012
\end{flushleft}

\end{titlepage}

\newpage
\tableofcontents
\newpage

\section{Introduction and Motivation}
\label{sec:intro}


String theory is a candidate for a fundamental theory of nature, providing at the same time a UV-finite quantum theory of gravity and unification of all forces and fermionic matter. Mathematical consistency requires string theory to live in a ten dimensional space-time, and a description of our large four-dimensional physics thus necessitates compactification of the additional six dimensions of space.

The need for compactification confronts us with two formidable consequences: Firstly, even given the known internal consistency constraints of string theory, there are unimaginably large numbers of 6d manifolds available for compactification. Secondly, many compact manifolds allow for continuous deformations of their size and shape while preserving their defining properties (such as topology, vanishing curvature, etc) -- these are the moduli, massless scalar fields in 4d. This moduli problem is exacerbated if we wish to arrange for low-energy supersymmetry in string theory, as compactifications particularly suitable for this job -- Calabi-Yau manifolds -- tend to come with hundreds of complex structure and K\"ahler moduli.

Therefore, a very basic requirement for string theory to make contact with low-energy physics is moduli stabilization -- the process of rendering the moduli fields very massive. Moreover, as supersymmetry is very obviously broken -- and so far has not been detected -- ideally, moduli stabilization should tolerate or even generate supersymmetry breaking. And finally, the process should produce a so-called meta-stable de Sitter (dS) vacuum with tiny positive cosmological constant, so as to accommodate the observational evidence for the accelerated expansion of our universe by dark energy~\cite{Perlmutter:1998np,Riess:1998cb,Komatsu:2010fb}.

The task of moduli stabilization and supersymmetry breaking has recently met with considerable progress, which is connected to the discovery of an enormous number~\cite{Bousso:2000xa,Giddings:2001yu,Kachru:2003aw,Susskind:2003kw,Douglas:2003um} of stable and meta-stable 4d vacua in string theory. The advent of this 'landscape'~\cite{Susskind:2003kw} of isolated, moduli stabilizing minima marks considerable progress in the formidable task of constructing realistic 4d string vacua.

A large aspect of these recent advances relies on the use of quantized closed string background fluxes in a given string compactification. These flux compactifications can stabilize the dilaton and the complex structure moduli of type IIB string theory compactified on a Calabi-Yau orientifold supersymmetrically~\cite{Giddings:2001yu}. The remaining volume moduli are then fixed supersymmetrically by non-perturbative effects, e.g. gaugino condensation on stacks of D7-branes~\cite{Kachru:2003aw}. The full effective action of such fluxed type IIB compactifications on Calabi-Yau orientifolds was derived in~\cite{Grimm:2004uq}. In type IIA string theory on a Calabi-Yau manifold all geometric moduli can be stabilized supersymmetrically by perturbative means using the larger set of fluxes available~\cite{DeWolfe:2005uu}.

Moduli stabilization itself is a necessary prerequisite for a successful description of cosmological inflation in string theory. For slow-roll inflation requires a separation of scales between the inflationary scalar degrees of freedom with masses lighter than the inflationary Hubble scale and every other scalar field which needs to be heavy. Beyond that, the slow-roll flatness of the scalar potential required during the last observable 60 e-folds of inflation requires a substantial amount of control over higher-dimension operators in the effective field theory. This further motivates embedding inflation into string theory as a UV-complete candidate for quantum gravity. 

Counting numbers of models, most successful inflationary model building in string theory has focused on the corner of the landscape described by type IIB flux compactifications on orientifolds of warped Calabi-Yau threefolds, and has produced small-field models of slow-roll inflation. A small-field model is characterized by the fact that the field range $\Delta\phi_{60}$ the inflaton traverses during the observable last 60 e-folds of inflation is less than $M_{\rm P}$. Planckian field traversal during inflation marks a critical boundary, as here one transitions from the need to control chiefly dimension-six operators in small-field inflation to the need to control correction to all orders in large-field inflation. The inflaton candidate in these constructions is often chosen to be the position of a mobile D-brane~\cite{Kachru:2003sx,Baumann:2007ah}, a combination of the geometric volume moduli of the Calabi-Yau~\cite{Conlon:2005jm,Bond:2006nc,Linde:2007jn,Cicoli:2008gp,Cicoli:2011ct}, or an axion originating in the higher $p$-form $NSNS$ and $RR$ gauge fields of string theory~\cite{BlancoPillado:2004ns,Kallosh:2007cc,Grimm:2007hs,Burgess:2008ir}.

There are several alternatives this plethora of slow-roll small-field models. One comes in the form of DBI inflation, where the specific form of the interactions of the inflaton dictated by the DBI action on a D-brane serve to slow down the field on a steep potential~\cite{Alishahiha:2004eh}. Another one consists of the idea of using a 'coherent' assistance effect of typically hundreds of string theory axions with sub-Planckian field range to yield an effective large-field model, called 'N-flation'~\cite{Dimopoulos:2005ac}. Finally, there are recent constructions harnessing monodromy of the potential energy sourced by branes or fluxes with respect to D-brane position or $p$-form axions. This monodromy inflation mechanism allows for parametrically large-field inflation in string theory driven by a single field~\cite{Silverstein:2008sg}. In the case of axion monodromy, the powerful shift symmetries of some of the $p$-form axions allow for a well-controlled and potentially large class of large-field models on warped type IIB Calabi-Yau compactifications~\cite{McAllister:2008hb,Flauger:2009ab,Berg:2009tg,Dong:2010in}.

For some recent reviews on flux compactifications and the associated questions of the landscape of meta-stable dS vacua and inflation in string theory, with a much more complete list of references, please see~\cite{Douglas:2006es,McAllister:2007bg,Baumann:2009ni}.

Observationally, the $\Delta\phi_{60}\sim M_{\rm P}$ boundary between small-field and large-field models has a second tantalizing aspect. This is the case because of the Lyth bound \cite{Lyth:1996im,Efstathiou:2005tq} shows that for any
single-field model of inflation, observable tensor modes in the CMB require a super-Planckian field range.
Upcoming CMB observations, both ground and space based, are projected to detect or constrain the tensor to scalar ratio $r$ at the level
$r\gtrsim 0.01$ (see for example \cite{Zaldarriaga:1996xe,Kamionkowski:1996ks,Efstathiou:2006ak,Efstathiou:2007gz,Bock:2006yf,Taylor:2006jw,Samtleben:2008rb,Pryke:2008xp,Baumann:2008aq,Efstathiou:2009xv,Bouchet:2009tr,Chiang:2009xsa,Sheehy:2011yf,Filippini:2011ds,Fraisse:2011xz}), while existing and upcoming satellite
experiments also significantly constrain the tilt of the spectrum \cite{Reichardt:2008ay,Komatsu:2010fb,Dunkley:2010ge}.

The tensor-to-scalar ratio $r$ is a single dimensionless number with a lot of discriminative power, separating single-field slow-roll inflation into two classes clearly distinguished by their dramatically different sensitivity to UV physics. Each of the two classes encompasses a large set of individually different models of inflation. Therefore, one may hope by summing over sizable samples of each class -- each sample averaging widely over many different model construction corners of the landscape -- and accounting for the influence of their respective population likelihood by cosmological dynamics, like tunneling, one can provide a statistical expectation as to whether $r=0$ or $r>0$ (in the observationally accessible sense of $r\gtrsim 0.01$). If successful, this would provide a second example where statistical reasoning on the landscape leads to a modest prediction akin to the weakly-anthropic explanation of the present-day small vacuum energy by scanning over $N_{vac.}\gtrsim 10^{500}$ flux vacua~\cite{Bousso:2000xa}.

The present work is an attempt to do so. We will show that an answer to this question can be reduced to the question of the number frequency distributions of small-field and large-field driving regions of the landscape, and of their vacuum energies -- that is, counting. Then we will fail, as we do not know how to count across the whole landscape, for its largest part is terra incognita still. A much more modest version of the counting task can be formulated for the region of the landscape described by flux-stabilized warped CY 3-fold compactifications of type IIB string theory in the description of F-theory on elliptically fibered 4-folds. A large sample (several millions) of such potentially low-energy supersymmetric compactifications are fully computationally accessible in terms of hypersurfaces in toric ambient space described completely the discrete data of the associated gauged linear-sigma models (GLSMs). We then outline for this sector of the landscape how to formulate the counting problem, and discuss the implications of a possible answer.  For this purpose, we start in Section~\ref{sec:input} with an analysis of those premises of possible arguments which follow from what is currently known about scalar fields in compactified string theory. Section~\ref{sec:argument} proceeds from these premises with an argument which essentially states that populating large-field models of inflation in string theory would cost exponentially dearly compared to seeding small-field models \emph{if} populated from the lowest-lying de Sitter (dS) vacuum of the whole landscape. Section~\ref{sec:counting} reviews the interplay of the dynamics of eternal inflation and tunneling as well as the anthropic requirements for a statistical explanation of the observed small cosmological constant (c.c.) to work. Its central outcome is the observation that the progenitor vacua of cosmologically and anthropically viable descendant regions of the landscape are of high-scale vacuum energy, and ultimately lead to democracy on the landscape. Hence the reduction to counting. Section~\ref{sec:discuss} then closes with a discussion of these arguments.

\section{Assumptions}
\label{sec:input}

Let us being by collecting some of the known results on scalar fields obtained from compactification of string theory to four dimensions. These results and properties of the types of scalar fields, the moduli, coming from a given compactification will form the premises of our later discussion of the prevalence of small-field versus large-field models of inflation in string theory.

\subsection{Need for symmetry}
\label{sec:symm}

We will start our walk through the premises by looking at the the need for a symmetry if large-field inflation driven by a single scalar field is to be embedded into string theory.
Large-field slow-roll inflation driven by a single scalar field is stable under radiative corrections only in presence of an effective symmetry suppressing higher-dimension operators to all orders, which must be dominantly broken by just the inflaton potential itself. This is immediately clear from the known argument to the effect of the so-called 'eta problem'. The classical background dynamics of single-field large-field inflation requires an inflaton potential dominated by a monomial
\beq
V_0(\phi)=\mu^{4-n}\phi^n\quad.
\eeq
Such a potential requires a minimum field range needed to generate the last observationally required $N_e\simeq 60$ e-folds of inflation of
\beq
\Delta\phi_{N_e}\simeq \sqrt{2n N_e}\,M_{\rm P}\gg M_{\rm P}\;\;\forall n\gg 0.01\quad.
\eeq
The minimal distance in field space traversed during the observationally accessible period of inflation is super-Planckian, that is 'large-field', in these models.

In absence of any further information constraining the effective field theory below $M_{\rm P}$ dimension-6 operators will be generated by quantum corrections. Among them we will generically operators of the type
\beq
\Delta V_6\sim V_0\,\frac{\phi^2}{M_{\rm P}^2}\quad.
\eeq
Such a correction will shift, in particular, the 2nd slow-roll parameter
\beq
\eta\equiv \frac{V''}{V}
\eeq
where $()'\equiv \partial/\partial\phi$ by a piece
\beq
\Delta\eta=\frac{\Delta V_6''}{V_0}\sim{\cal O}(1)
\eeq
which destroys slow-roll inflation as soon as the inflaton moves about a Planck unit.

The only known way to forbid these dangerous higher-dimension operators to all orders in perturbation theory for an elementary scalar field (besides supersymmetry or conformal symmetry which, however, are incompatible with positive vacuum energy) is a shift symmetry. An unbroken shift symmetry requires a constant scalar potential. Assume now, that the dominant source of soft breaking of the shift symmetry is the field-dependence of the scalar potential itself, and its potential energy density is sufficiently below the cutoff of the effective field theory. Then radiative corrections induced by this inflationary soft shift symmetry breaking are proportional to powers of the order parameter of the symmetry breaking, namely
\beq
\frac{V_0(\phi)}{M_{\rm P}^4}\quad.
\eeq
So they become large only at extremely large super-Planckian field values if $V_0(\phi_{60})\ll M_{\rm P}^4$. This is the original idea of large-field chaotic inflation~\cite{Linde:1983gd}.

Parametrically large-field inflation with a single scalar field in string theory will thus constrain us to finding candidate scalar field protected by a very good shift symmetry. This essentially limits us to the axions of string theory, as we will see.

\subsection{Properties of scalar fields in compactified string theory}
\label{sec:scalars}

Let us now look at the general properties of scalar fields arising from compactification of string theory to four dimensions.\footnote{We do neglect the dilaton here despite it being there already in 10d because by definition it couples to everything and cannot be a good large-field inflation candidate.} These consist of the geometrical closed string moduli related to massless deformations of the internal compact manifold, open string moduli such as transverse positions of D-branes mutually supersymmetric and BPS with respect to the background manifold, and pseudoscalar 'axions' from the NSNS and RR $p$-form gauge fields.

The geometrical closed string moduli, such as volumes or shapes of the internal manifold, are not protected by fundamental shift symmetries. 'Fundamental' is used here in the sense, that it describes a shift symmetry of a pseudoscalar field which is inherited from a gauge symmetry already present on the worldsheet. Geometric moduli therefore are generically not useful for parametrically large-field behaviour, although they allow very well for a plethora of small-field string inflation models.\footnote{In some cases, the structure of string compactifications on Calabi-Yau allows for a so-called 'extended no-scale structure' of the Kahler potential of the volume moduli which at large volume leads to suppression of the otherwise large 1-loop string corrections. In such cases, this enables the use of the certain volume moduli to generate an inflaton potential of the type
$$ V_0(\phi)\sim 1-e^{-\alpha \phi}$$ which resides at the borderline between small-field and large-field models (e.g. fibre inflation in type IIB~\cite{Cicoli:2008gp}).}

Open string moduli, i.e. the position of the D-branes, have been shown in general to possess a sub-Planckian field-range for the canonically normalized scalar field corresponding to a given D-brane position modulus. Canonical normalization of the open string moduli involves inverse powers of the size of internal manifold which for control reasons must be always larger than the string length $\sqrt{\alpha'}$~\cite{Baumann:2006cd}. Therefore
\beq
\Delta\phi_{\rm D-brane\,pos.}\sim M_{\rm P} \left(\frac{\sqrt{\alpha'}}{R}\right)^p \lesssim M_{\rm P}\quad,\quad p\geq1
\eeq
cannot be parametrically super-Planckian. Besides that, open string moduli generally do not inherit good shift symmetries. A similar parametric argument on the field range holds for the geometrical closed string moduli as well.

The remaining class of scalar fields from compactification are axionic fields which arises from integration of the NSNS 2-form gauge field $B_2$ or RR $p$-form gauge fields $C_p$ over cycles of the internal manifold. This gives rise to axion fields
\beq b_i=\int_{\Sigma_2^i}B_2 \qquad,\qquad c_\alpha^{(p)}=\int_{\Sigma_p^\alpha}C_p
\eeq
where $\Sigma_2^i$ denotes $i^{\rm th}$ 2-cycle, and $\Sigma_p^\alpha$ denotes the $\alpha^{\rm th}$ $p$-cycle. The gauge symmetries of the $p$-form gauge fields from the worldsheet translate into shift symmetries of the dual pseudoscalar 4d axion fields. These continuous shift symmetries are broken to a discrete subgroup by instanton effects, and sometimes also by the effects of orientifold projections introduced to further break supersymmetry. An example of the latter is type IIB compactified on an O7 orientifolded Calabi-Yau manifold with fluxes and D3-branes, where the orientifolding breaks the shift symmetry of axions coming from $B_2$~\cite{Silverstein:2008sg,McAllister:2008hb,Flauger:2009ab,Berg:2009tg,Dong:2010in}. The $p$-form induced axions in string theory, denoted collectively with $a_I$,  are therefore periodic on fundamental domain with limited field range~\cite{Banks:2003sx,Svrcek:2006yi,Grimm:2007hs}
\beq \Delta a_I=(2\pi)^2\quad.
\eeq
Converting this into canonically normalized scalar fields involves again inverse powers of the size of the internal manifold. The result behaves similar to the case of the open string moduli~\cite{Baumann:2006cd}
\beq
\Delta\phi_{\rm axions}\sim M_{\rm P} \left(\frac{\sqrt{\alpha'}}{R}\right)^p \Delta a_I \lesssim M_{\rm P}\quad,\quad p\geq1\quad.
\eeq

These results have lead to the notion of no-go statement: there are no scalar fields with an intrinsically super-Planckian field range coming out of 4d string compactifications~\cite{Banks:2003sx,Svrcek:2006yi,Baumann:2006cd,Grimm:2007hs}.

\subsection{Populating vacua -- tunneling and quantum diffusion}
\label{sec:tunnel}

The final premise we need to discuss concerns the mechanism to populate meta-stable a given dS vacuum, and by extension a string theoretic inflationary region, in the landscape. The notion of vacua as the (meta)stable ground states of a local QFT as an effective field theory derived from string theory exists only the regime of controlled 4d low energy approximation to string theory. This is realized only in the large-volume and weak string coupling regime, when the supergravity approximation supplemented by the leading string-loop, $\alpha'$- and non-perturbative corrections is valid. Within this region of controlled approximations the only known mechanism of vacuum transitions at zero temperature proceeds via field theoretic tunneling.

There are basically two known Euclidean instantons describing tunneling in QFT, and one process based on the quantum fluctuations of a light scalar field in dS space. The string landscape consists of a moduli scalar potential for an ${\cal O}(100\ldots 1000)$-dimensional scalar field space which has upwards of ${\cal O}(10^{500})$ isolated local minima. The local situation of tunneling between two adjacent vacua in moduli space is the often described by a scalar potential $V(\chi_i)$ of the canonically normalized moduli $\chi_i$ which has two local minima $\chi_{i,\pm}$ separated by a finite potential barrier. Let us call $\chi_{i,+}$ and $\chi_{i,-}$ the false, and true vacuum, respectively. These minima are separated by generically sub-Planckian distances in field space $$|\Delta\chi_{i,\pm}|=\sqrt{\sum_i (\chi_{i,+}-\chi_{i,-})^2}\lesssim M_{\rm P}\quad.$$

If in this general situation the barrier height is non-negligible compared the vacuum energy difference $\Delta V=V(\chi_{i,+})-V(\chi_{i,-})$, then the dominant Euclidean instanton contributing to tunneling is the Coleman-DeLuccia (CDL) instanton~\cite{Coleman:1977py,Coleman:1980aw}. In flat space this instanton is described by the so-called $SO(4)$ symmetric 'bounce solution' to the Euclidean field equations in the 'inverted' scalar potential $-V(\chi_i)$
\beq
\label{bounceEOM}
\frac{d^2\chi_i}{d\rho^2}+\frac{3}{\rho}\,\frac{d\chi_i}{d\rho}=\frac{\partial V}{\partial \chi_i}\quad\forall i\quad.
\eeq
Here $\rho=\sqrt{\tau^2+|\vec r|^2}$ denotes the $SO(4)$ symmetric radial variable with $\tau=it$ Euclidean time. The boundary conditions on the bounce solution require
\beq
\chi_i(\tau=0,\vec r)\;=\;\chi_{i,0}\simeq \chi_{i,-}\quad,\quad \chi(\tau=0,\vec r)\;\xrightarrow[|\vec r|\to\infty]{}\;\chi_{i,+}\quad,\quad \partial_\tau\chi_i(\tau=0,\vec r)\;=\;0\quad.
\eeq
In terms of $\rho$, $\chi_i(\rho)$ they read as
\beq
\label{bounceICS}
\chi_i(0)\;=\;\chi_{i,0}\simeq\chi_{i,-}\quad,\quad \chi_i(\rho)\;\xrightarrow[\rho\to\infty]{}\;\chi_{i,+}\quad,\quad \left.\frac{d\chi_i}{d\rho}\right|_{\rho=0}\;=\;0\quad.
\eeq
One then computes the Euclidean action on the bounce solution, called here $\chi_i^{(b)}(\rho)$,
\beq
S_E[\chi_i^{(b)}]=-\int d\rho\, \rho^3 \left[\frac12 \left(\frac{d\vec \chi}{d\rho}\right)^2+V(\chi_i)\right]
\eeq
and the tunneling rate is given by
\beq
\Gamma_{CDL}\sim e^{-B} \quad,\quad B=S_E[\chi_i^{(b)}]-S_E[\chi_{i,+}]\quad.
\eeq

The presence of gravity yields corrections to this result which become important in two situations. For one, this happens if the barrier thickness becomes super-Planckian, which is generically not the case for next-neighbour local minima in the landscape moduli potential. The other situation occurs when the false vacuum approaches zero vacuum energy with the true vacuum being AdS. Then gravitational suppression of tunneling can happen. This case is irrelevant for our situation of having to populate a possible inflationary region of the landscape from a prior, higher-lying nearby dS vacuum. Therefore, CDL tunneling in our dS-to-dS situation with sub-Planckian barrier thickness can be described by flat space CDL tunneling neglecting gravity.

This generic picture of a landscape populated CDL tunneling has been studied extensively in the literature. 
In certain controlled constructions such as warped Calabi-Yau compactifications of type IIB string theory with imaginary self-dual 3-form fluxes~\cite{Giddings:2001yu} and anti-D3-branes provide a string theoretic realization of the effective CDL tunneling description in terms of the derived moduli potential~\cite{Kachru:2002gs,Freivogel:2008wm}.

The exception to this situation is the case where the barrier becomes very shallow and flat. In that case the gravitational corrections to tunneling become very important, and the Euclidean solution with the smallest action $B$ mediating tunneling switches to the Hawking-Moss instanton~\cite{Hawking:1981fz}. This describes tunneling from the false vacuum at $\chi_{i,+}$ to the barrier top at $\chi_{i,T}$ with a rate
\beq
\label{HMProb}
\Gamma_{HM}\sim e^{-M_{\rm P}^4\left(\frac{1}{V(\chi_{i,+})}-\frac{1}{V(\chi_{i,T})}\right)}\quad.
\eeq

Finally, one can show that this process is essentially equivalent to a description where the light scalar fields $\chi_i$ close to the false dS vacuum at $\chi_{i,+}$ are driven by the dS quantum fluctuations up the barrier onto its top. The magnitude of the dS quantum fluctuations which form a Gaussian random field are given by (see e.g.~\cite{Linde:2005ht})
\beq
\label{dSvariance}
\langle \chi_i^2\rangle=\frac{3 H_+^4}{8\pi^2 m_i^2}
\eeq
where $H_+=\sqrt{V(\chi_{i,+})}$ denotes the Hubble constant of the false dS vacuum. The diffusion probability to reach the barrier top
\beq
\label{diffProb}
\Gamma_{diff}\sim e^{-\frac{\sum_i (\chi_{i,T}-\chi_{i,+})^2}{\sum_i \langle \chi_i^2\rangle}}\sim \Gamma_{HM}
\eeq
behaves like the one derived from the Hawking-Moss instanton.

For all processes it is visible that a possible minimal field displacement will be expensive in terms of tunneling rate suppression if such a displacement due to tunneling were required by the initial conditions of a given inflation model.

\section{An almost argument -- field displacement is expensive}
\label{sec:argument}


We will now explore the consequences of the premises outlined in the last section.

We know from the discussion in subsection~\ref{sec:symm} that large-field inflation driven by a fundamental scalar field needs an effective shift symmetry to be radiatively stable. The only other known mechanism for curing the radiative instability of generic scalar field theories, supersymmetry, forbids positive vacuum energy which renders it useless for inflationary purposes. From subsection~\ref{sec:scalars} we have that essentially all scalar fields from compactifying string theory to four dimensions will have a sub-Planckian, or in some case just-so Planckian, intrinsic field range.

Now consider that, in particular, the fundamental domain of all scalar fields with good shift symmetries from string theory, the p-form axions, is limited. Then by the very definition of a fundamental domain the only way beyond this point consists of finding an effect which unwraps the fundamental domain onto its covering space. This is called monodromy. For inflationary purposes this indicates two things. Firstly, the required effect has be something which gives potential energy to the candidate inflaton field in a given string compactification. Secondly, it must possess monodromy in the inflaton in order to see its covering space instead of its limited fundamental domain.

This would result in parametrical large-field inflation (kinematically at least, up to back reaction constraints). Some form of non-trivial monodromy of the potential energy of the candidate inflaton with respect that very same inflaton is thus necessary for large-field inflation. We see, that this follows from the very definition of the limitation of the field range for those fields, the axions, which possess potentially good shift symmetries.

As parametrical large-field inflation in string theory can only proceed given a good shift symmetry, the only viable candidates are the $p$-form axions and we are left with trying to generate axion monodromy in some form of potential energy for the axion. A potential energy for the $p$-form axions -- which spontaneously breaks their shift symmetry -- is generated by instanton effects, branes or fluxes. Monodromy in the potential energy with respect to the $p$-form axions exists for $(p+3)$-branes wrapped on $p$-cycles~\cite{McAllister:2008hb}, or non-topological fluxes involving the $p$-form gauge field on a $p$-cycle~\cite{Dong:2010in}. The crucial aspect here is the fact that the monodromy-carrying objects, the branes and non-topological fluxes, spontaneously break the axionic shift symmetry the same way they also spontaneously break supersymmetry in a regime controlled typically by warping. Therefore, parametrical large-field inflation in string theory will come from some form of axion monodromy.

For the case of axions from $c=\int_{\Sigma_2}C_2$ in type IIB on a homologous pair of 2-cycles in a pair of warped throats wrapped by an $NS5\overline{NS5}$-brane pair~\cite{McAllister:2008hb} this leads to a large-field potential
\beq
V_0(\phi)\simeq \mu^3\sqrt{{\rm vol}_{\Sigma_2}^2+\phi^2}\quad{\rm with}\quad \phi\sim c\quad.
\eeq
The back reaction of the inflationary vacuum energy stored in the wound-up axion on the moduli potential as well as differing types of potential energy with axion monodromy such as other branes or fluxes will generically lead to a flattening-out of the axion inflaton potential~\cite{Dong:2010in}. In general we expect large-field potentials from axion monodromy to behave like
\beq
V_0(\phi)\sim\left\{\begin{array}{c}\phi^2\quad{\rm for}\quad \phi\lesssim M_{\rm P}   \\ \phi^p\;\;,p\lesssim 1\quad{\rm for}\quad \phi\gg M_{\rm P}\end{array}\right. \quad.
\eeq

There is a crucial observation to be made here. The scalar potential during axion monodromy inflation is driving both the leading shift symmetry breaking and the back reaction. Therefore, the point of vanishing axion vev being also the point of vanishing axion-induced $D3$-brane charge and potential energy is a minimum of the inflaton potential. This holds to high accuracy independently of the shape or structure of the moduli potential, or the back reaction of the compactification geometry. If it were otherwise, the non-universality of the back reaction from the moduli potential or the geometry would destroy the shift symmetry in a non-universal way to begin with. This, by construction, cannot happen as the sole source of back reaction is controlled by the same parametrically weak effect which spontaneously breaks the shift symmetry to leading order in the first place.

Let us denote the moduli sector other than the inflaton axion collectively with field(s) $\chi$. Then we can write the full scalar including both moduli and large-field inflation from string theory following from the premises of section~\ref{sec:input} and the discussion above as
\beq
V(\phi,\chi)=V_0(\phi)+U_{mod.}(\chi)\quad.
\eeq

The final aspect of our discussion below will involve the dynamics of populating a potential large-field inflation model in string theory. The only way known to exist proceeds via quantum tunneling from a prior meta-stable dS vacuum (see subsection~\ref{sec:tunnel}). Thus, we will only need to discuss the immediate neighbourhood of the moduli potential with respect to our current vacuum. Tunneling will proceed dominantly from the closest-by higher-lying dS minimum of the moduli potential with the smallest barrier height at the point where the Euclidean bounce solution crosses the barrier. Again, this relies on the decoupling between the position of the minimum of the inflaton potential and the moduli potential due the strong axionic shift symmetry. Neglecting the generally curved trajectory of the multi-field bounce solution, we will take the moduli potential as approximated by a '1d section along the bounce'. Then the existence of two close-by non-degenerate dS minima can be modeled by a quartic polynomial
\beq
\label{UmodApprox}
U_{mod.}(\chi)=\lambda\,\chi^4+g\,\chi^3+m^2\,\chi^2\quad.
\eeq
Such a local neighbourhood structure of the moduli potential relies on two properties. Firstly, the string landscape admits an extremely large number of isolated minima of the moduli potential. Second, the intrinsic field range limitation of all moduli fields implies that most of these minima must have distance $\ll M_{\rm P}$. This justifies Taylor expanding the potential around a given minimum towards the closest neighbour.

At last, there are subleading sources of shift symmetry breaking, typically non-perturbative effects. The presence of these instanton effects generates a periodic potential for the $p$-form axions. Its period is given by the extent of their fundamental domain, $2\pi f$ in terms of the axion decay constant $f$. The magnitude of the instanton-induced axion potential is exponentially suppressed at large volumes~\cite{Witten:1996bn,McAllister:2008hb}. This gives us
\beq
\delta V_{non-pert.}(\phi)=\Lambda^4\,\cos\left(\frac{\phi}{2\pi f}\right)\quad,\quad\Lambda^4\sim e^{-2\pi\cdot{\rm vol}_\Sigma}
\eeq
where ${\rm vol}_\Sigma$ denotes the volume of the cycle threaded by the axion. We can thus easily dial them negligibly small. Yet their presence will be crucial for the relative count of small-field inflation models in string theory compared to the axion monodromy based large-field models.

\begin{figure}[t]
\begin{center}
\includegraphics[width=0.95\textwidth]{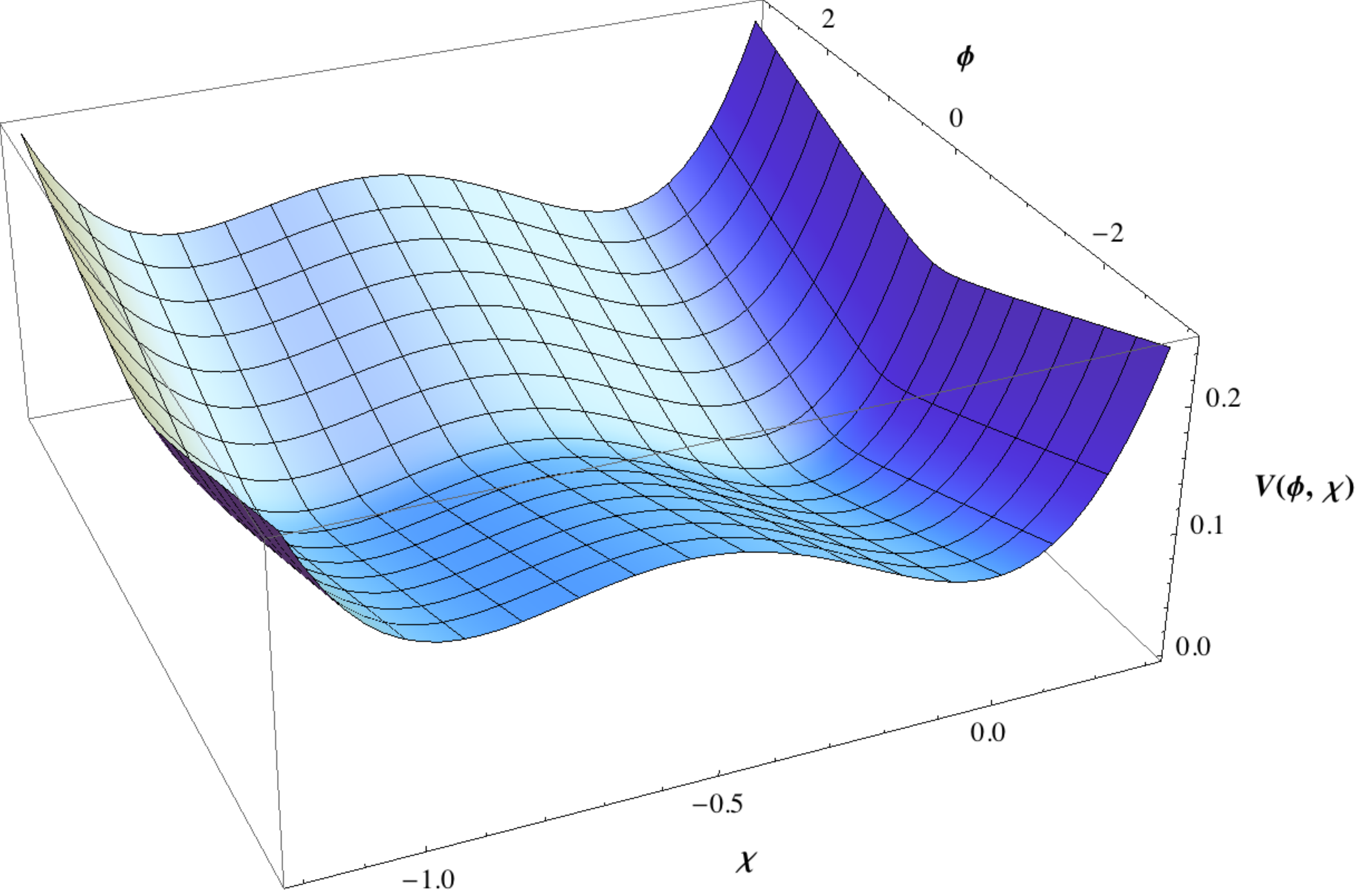}
\end{center}
\refstepcounter{figure}\label{Vplot1}
\vspace*{-.2cm} {\bf Figure~\ref{Vplot1}:} The potential $V(\phi,\chi)$ in arbitrary units for
the canonically normalized inflaton field $\phi$ and modulus $\chi$ (in Planck units) in the large-field regime $b\ll 1$.  The distance in field space at about $\phi\simeq 11\,M_{\rm
P}$ is the minimal field range necessary to get 60 e-folds of slow-roll inflation.
\end{figure}

The full scalar potential for large-field inflation in string theory in presence of the local moduli potential, under the premises of section~\ref{sec:input}, thus reads
\beq
V(\phi,\chi)=V_0(\phi)+\delta V_{non-pert.}(\phi)+U_{mod.}(\chi)\quad.
\eeq
We will consider for the sake of explicitness $V_0(\phi)$ of $C_2$ axion monodromy from $NS5$-branes, but the conclusions drawn from here will be general and do not depend on the precise choice $V_0(\phi)$
\beq
\label{Vtot}
V(\phi,\chi)= \mu^3\left[\sqrt{{\rm vol}_{\Sigma_2}^2+\phi^2}+bf\,\cos\left(\frac{\phi}{2\pi f}\right)\right]+U_{mod.}(\chi)\quad.
\eeq
This form introduces the slope parameter
\beq
b=\frac{\Lambda^4}{\mu^3 f}
\eeq
True large-field behaviour requires $b\ll 1$, or equivalently ${\rm vol}_{\Sigma_2}\gg 1$. For this case, the example potential is shown in Fig.~\ref{Vplot1}. Let us label again the two adjacent (meta)stable minima of the moduli potential by $\chi_\pm$ with $\chi_+<\chi_-$ and place $\chi_-=0$.

Conversely, fine-tuning $b$ around $b\simeq 1$ will have near-flat inflection points appearing in the potential~\cite{Flauger:2009ab}. As $f<M_{\rm P}$, these inflection will be spaced with sub-Planckian distances. Therefore, moderately fine-tuning one of them to slow-roll flatness around the inflection point by using $b$ will result in small-field inflation.

This leads to a crucial result: Under the premises of section~\ref{sec:input}, there will be at least one model of small-field inflation contained in \emph{every} working model of large-field inflation in string theory.

The small-field and large-field parameter regions of axion monodromy occupy different volumes of microscopic parameter space, as the occurrence of a slow-roll flat inflection point needs a significant tuning in $b\simeq 1$ in terms of the microscopic parameters, such as fluxes. Such fine-tuning can range from moderate ${\cal O}(10^{-2})$~\cite{BlancoPillado:2004ns,Baumann:2007ah} to more severe values of ${\cal O}(10^{-8})$~\cite{Linde:2007jn}. However, the number frequency hierarchy deriving from the tune is finite, and can be easily dominated by exponential ratios from the dynamics of populating all these different models, which is tunneling.

\begin{figure}[t]
\begin{center}
\includegraphics[width=0.95\textwidth]{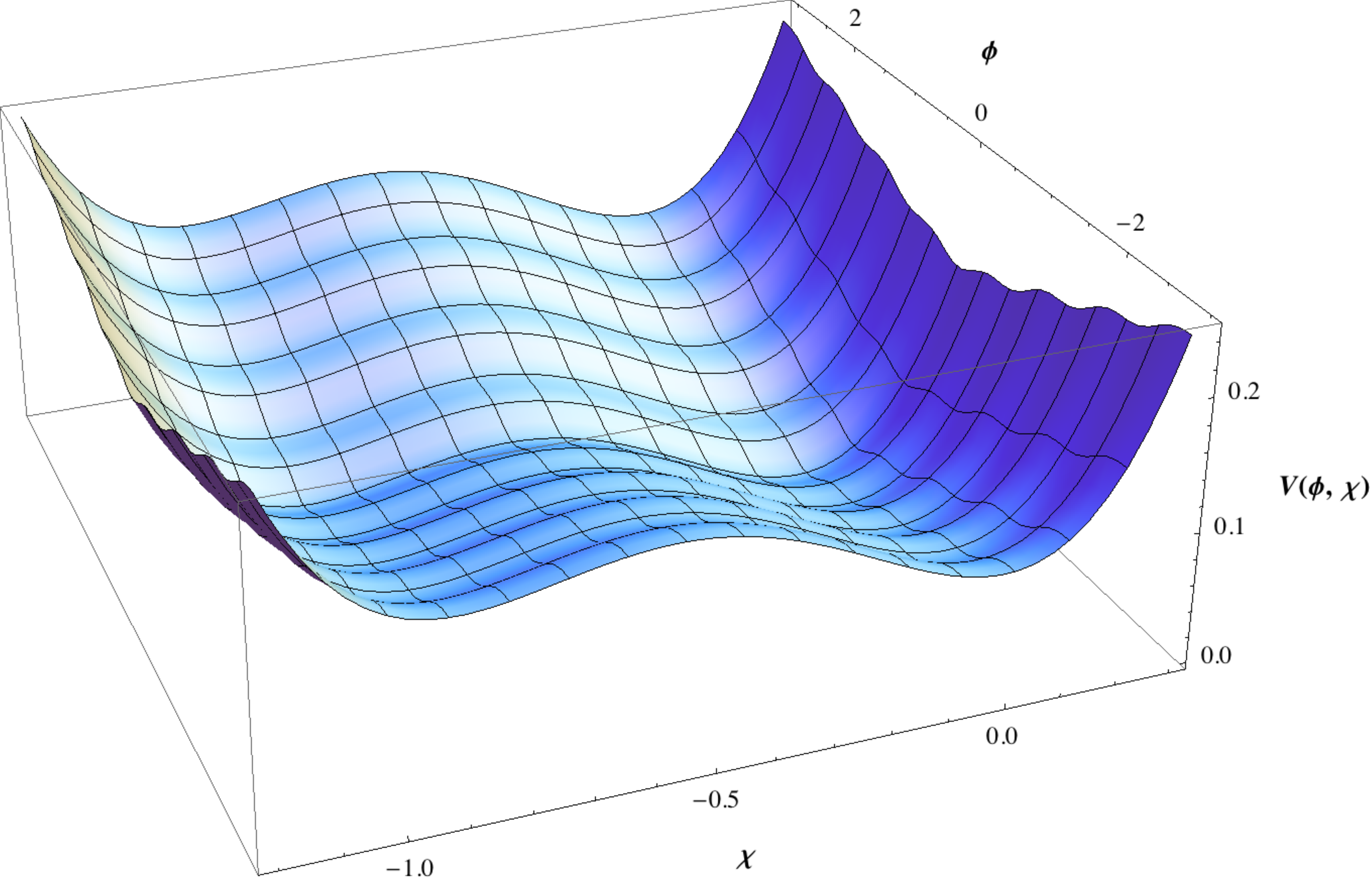}
\end{center}
\refstepcounter{figure}\label{Vplot2}
\vspace*{-.2cm} {\bf Figure~\ref{Vplot2}:} The potential $V(\phi,\chi)$ in arbitrary units for
the canonically normalized inflaton field $\phi$ and modulus $\chi$ (in Planck units) in the small-field regime $b\simeq 1$.  The distance in field space at about $\phi\simeq 11\,M_{\rm
P}$ is the minimal field range necessary to get 60 e-folds of slow-roll inflation.
\end{figure}

We will now take a look at the process of tunneling into the inflationary valley of $\chi_-=0$ from the close-by valley at $\chi_+<0$. Behind this is the premise of subsection~\ref{sec:tunnel} that tunneling is the only process for cold vacuum transitions in the landscape which is known. For this purpose, one more property of the full large-field scalar potential eq.~\eqref{Vtot} is crucial. Generically, it is the instanton-induced inflection point closest to post-inflationary minimum at $\chi_-=\phi_-=0$ which is the only one suitable for small-field inflation. The reason is the upper bound on $f$. For typical values of $f$ several of the inflection points will sit within the quadratic region $\phi\lesssim M_{\rm P}$ close to the origin of $V_0(\phi)$. Unless the lowest-lying, and thus closest-to-origin inflection point is tuned inflationary flat using $b\simeq 1$, then there will be local minima at lower-lying inflection points which would trap the inflaton in false vacua. Conversely, the higher-lying inflection points above the fine-tuned one are too steep to support small-field inflation. An example of such an instanton-induced small-field contained in every stringy large-field model is shown in Fig.~\ref{Vplot2}. The inflationary inflection points are the two ones closest to $\phi=0$.

We will first look at tunneling mediated by the CDL instanton in the flat space approximation~\cite{Coleman:1977py,Coleman:1980aw}. As discussed in subsection~\ref{sec:tunnel}, the gravitational correction factor from including gravity is typically not important for tunneling from one dS to another lower dS vacuum. 

Let us first discuss which locale in the false-vacuum valley close to $\phi_+=0$ we expect to be the most likely starting position for any tunneling process. We may a priori expect the false valley to get populated by an even earlier tunneling event. That event may exit in particular at some $\phi\neq 0$ up the valley from where a further tunneling could start. However, the false valley is slow-roll in $\phi$ even more than the true valley close to $\chi_-=0$, and it supports false-vacuum eternal inflation at its false minimum $\chi_+<0,\phi_+=0$. Therefore, the ambient space-time residing initially in the false valley will have a fraction exponentially close to unity which actually sits at the false vacuum. The false vacuum inside the false valley thus exponentially dominates the initial state for any subsequent tunneling towards the true valley.

Next, getting slow-roll inflation in the true valley after tunneling there places a constraint on the exit state directly after tunneling. We must have the position of $\phi$ after tunneling, called $\phi_0$, supporting at least 60 e-folds of slow-roll inflation in the true valley
\beq
|\phi_0|\geq \phi_{60}\quad.
\eeq
For the large-field case $b\ll1$ this implies $$|\phi_0|\gg M_{\rm P}\quad,$$while in the small-field situation with $b\simeq 1$ we have $$0<\phi_{60}\leq|\phi_0|\lesssim M_{\rm P}\quad.$$
Note, that the fields may exit at $\chi_0<0$ and $|\phi_0|\gg\phi_{60}$ and possess finite speeds $\phi'(\rho),\chi'(\rho)$ as well, even in the small-field case. This is due the the negative spatial curvature inside a freshly formed CDL bubble. Negative curvature will serve to slow scalar fields regardless of their initial conditions enough to track them into slow-roll even on a small-field inflection point~\cite{Freivogel:2005vv,Dutta:2011fe}~\footnote{For recent discussions of other aspects of post-tunneling 'open inflation'~\cite{Linde:1998iw}, see~\cite{Yamauchi:2011qq,Sugimura:2011tk}.}. A viable bounce thus requires boundary conditions according to eq.~\eqref{bounceICS}
\beq
\label{bounceICS1}
\phi\xrightarrow[\rho\to\infty]{}\phi_+=0\quad,\quad\chi\xrightarrow[\rho\to\infty]{}\chi_+<0\quad,\quad|\phi(0)|=\phi_0>\phi_{60}\quad,\quad\chi_+\ll\chi(0)=\chi_0<0
\eeq
and for reasons of regularity also
\beq
\label{bounceICS2}
\left.\frac{d\phi}{d\rho}\right|_{\rho=0}=0\quad,\quad\left.\frac{d\chi}{d\rho}\right|_{\rho=0}=0\quad.
\eeq
There is also an energetics constraint which reads
\beq
V_+\equiv V(\phi_+,\chi_+)>V_-\equiv V(\phi_-,\chi_{60})
\eeq
to get down tunneling. 

If we look now at the inverted potential $-V(\phi,\chi)$ driving the Euclidean dynamics, eq.~\eqref{bounceEOM}, we see that such a bounce which originates downhill from the ridge and ends uphill on the ridge at $\chi_+<0,\phi_+=0$ does not exist. The gradient of $-V(\phi,\chi)$ always points away from $\phi=0$. Even allowing finite initial speed $\phi'(0),\chi'(0)$, neglecting the regularity boundary conditions eq.~\eqref{bounceICS2}, does not avoid this fate, as the friction term in the bounce e.o.m. eq.~\eqref{bounceEOM} immediately destroys any initial speed~\cite{Dutta:2011fe}.

This forces us to consider processes which move the field $\phi$ \emph{uphill} in the false valley to values $|\phi|>\phi_0>\phi_{60}$. A CDL bounce ending here and thus curving \emph{downhill} (in the inverted potential $-V(\phi,\chi)$) from its starting point $\phi_0,\chi_0$ is then possible \emph{after} $\phi$ has moved uphill (in $V(\phi,\chi)$) in the false valley.

How do we move up the false valley? The false vacuum at $\chi_+<0,\phi_+=0$ drives false-vacuum eternal inflation. As $m_\phi<H$, $\phi$ undergoes scale-invariant dS quantum fluctuations. These can be thought of as $\phi$ performing a Gaussian random walk with variance eq.~\eqref{dSvariance}
\beq
\langle \phi^2\rangle=\frac{3H_+^4}{8\pi^2m_\phi^2}\quad.
\eeq
This quantum diffusion has small probability for large jumps $\Delta\phi$ given in eq.~\eqref{diffProb}
\beq
P(\Delta\phi)\sim e^{-\frac{8\pi^2m_\phi^2\Delta\phi^2}{3H_+^4}}\quad.
\eeq
As $m_\phi\lesssim 10^{-5}\,M_{\rm P}$ from the inflationary constraints in the true valley, we see that
\beq
\frac{8\pi^2m_\phi^2}{3H_+^4}M_{\rm P}^2\gtrsim 1
\eeq
as long as $V(\phi_+,\chi_+)^{1/4}\lesssim 0.1\,M_{\rm P}$.

Therefore we arrive at a combination of two results. For \emph{every} large-field inflation model in string theory there exists at least one small-field model contained within via moderate tuning. However, the relative probability of realizing them dynamically via first quantum diffusing uphill in the false valley and then CDL tunneling out, is given by
\beq
\label{DOOM}
\frac{P(\Delta\phi_{large-field})}{P(\Delta\phi_{small-field})}\lesssim \frac{e^{-\frac{\Delta\phi_{large-field}^2}{M_{\rm P}^2}}}{e^{-\frac{\Delta\phi_{small-field}^2}{M_{\rm P}^2}}}\sim e^{-2 p N_e}\ll 1
\eeq
for large-field models $V_0(\phi)\sim \phi^p$, $p>1/(2N_e)$. Here we have used the large-field model $N_e$ e-folds interval
\beq
\phi_{N_e}\simeq \sqrt{2pN_e}
\eeq
and typically $N_e\simeq 60$ observationally.

We can check this result by replacing the quasi-rectangular path just considered by looking at quantum diffusion directly from the false vacuum at $(\phi_+,\chi_+)$ to a point $(|\phi|>\phi_{60}\,,\,\chi_T$ on the top of the potential barrier ridge at $\chi_T$ (which is then followed by classical rolling into the true valley). This process, as discussed in subsection~\ref{sec:tunnel}, is equivalent to a Hawking-Moss instanton~\cite{Hawking:1981fz} which tunnels to said point on the barrier ridge. Both descriptions yield parametrically the same result as found above in eq.~\eqref{DOOM}.

\begin{figure}[ht]
\begin{center}
\includegraphics[width=0.91\textwidth]{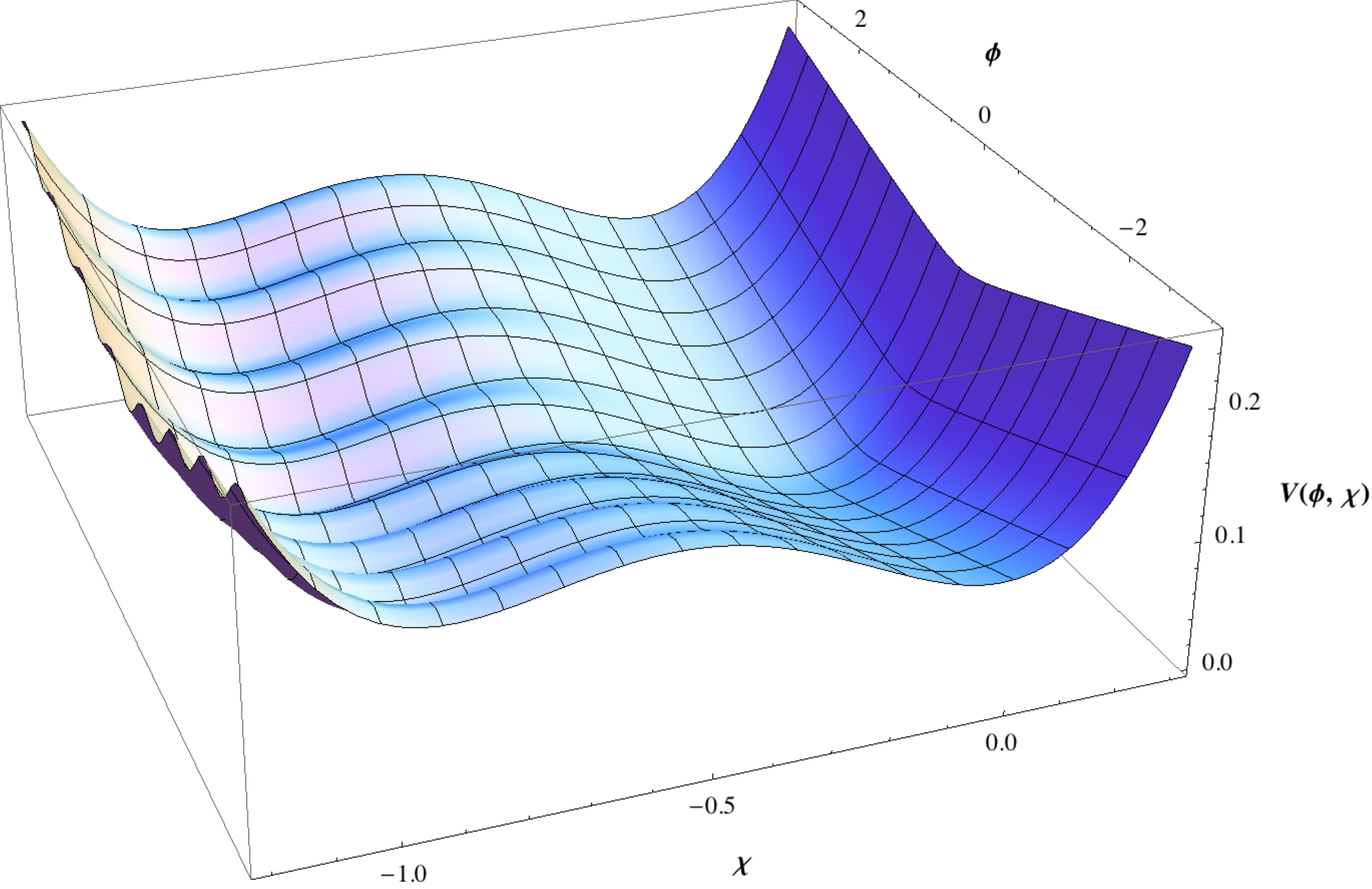}
\end{center}
\refstepcounter{figure}\label{Vplot3}
\vspace*{-.2cm} {\bf Figure~\ref{Vplot3}:} A generic potential $V(\phi,\chi)$ in arbitrary units for
the canonically normalized inflaton field $\phi$ and modulus $\chi$ (in Planck units). The magnitude of the instanton corrections $b\simeq 1$ typically varies over moduli space. A situation as depicted, where the false-vacuum inflaton valley has repeating local dS minima from the instanton effects, while across a distance in moduli space the true-vacuum inflaton valley may is smoothly in the large-field regime, is generic in the landscape.  The distance in field space at about $\phi\simeq 11\,M_{\rm
P}$ is the minimal field range necessary to get 60 e-folds of slow-roll inflation in the large-field regime of this example.
\end{figure}

This exponential advantage of the instanton-induced small-field regime crucially depends on the the necessity of uphill tunneling in the false-vacuum valley prior to the barrier-traversing tunneling.  Uphill tunneling in the false valley to provide a starting point for the barrier-crossing tunneling in moduli space is necessary only if the magnitude of the instanton correction is constant over moduli space. However, this assumption 
\beq
b=const.
\eeq
in eq.~\eqref{Vtot} does \emph{not} follow from the premises of section~\ref{sec:input}. Indeed, the very arguments used above which imply the shift symmetry of the inflaton axion forcing a decoupling between the large-field potential $V_0(\phi)$ and the moduli potential $U_{mod.}(\chi)$, imply also that $b$ can vary over moduli space. The instanton corrections break the continuous shift symmetry to a discrete subgroup. This fixes the argument of their sinusoidal dependence completely. In particular, the decoupling argument from the shift symmetry dictates that any phase $\varphi$ in the sinusoidal dependence of the instanton correction
\beq
\delta V_{non-pert.}\sim b f\cos\left(\frac{\phi}{2\pi f}+\varphi\right)
\eeq
is constant over moduli space to a high degree
\beq
\varphi=const. \neq \varphi(\chi)\quad.
\eeq
However, its magnitude can and will generically vary widely over moduli space. This happens because changing $b$ will preserve the positions of the critical points of the instanton correction which is what the discrete remainder of the shift symmetry demands
\beq
b=b(\chi)\neq const.\quad.
\eeq
An example of a potential from the axion monodromy large-field mechanism of string theory, which represents this generic situation on the landscape is depicted in Fig.~\ref{Vplot3}.

This changes the picture in one crucial aspect. Tunneling in moduli space can now easily provide for access to the true-vacuum valley of slow-roll inflation, where $b\lesssim 1$, from a false-vacuum valley with $b> 1$, as $b$ will generically change upon barrier traversal. Therefore, in the false-vacuum valley there will be generically often a wash-board potential for $\phi$ from the repeating multiple minima at $b > 1$. Because of $f< M_{\rm P}$ this will typically provide for false vacua close enough to any $\phi_{60}$ needed, regardless whether $\phi_{60}\lesssim M_{\rm P}$ or $\phi_{60}\gg M_{\rm P}$.\footnote{A related situation, where tunneling from a false vacuum $(\phi_+,\chi_+)$ with $\phi_+\simeq \phi_{60}$ and a quadratic potential steep in $\phi$ centered around $\phi_+$ which becomes a shallow quadratic potential centered around $\phi_-=0$ after tunneling in the $\chi$-direction, was recently studied as a simple toy model of $m^2\phi^2$ open chaotic inflation~\cite{Sugimura:2011tk}.} These false vacua (naively) are on equal footing\footnote{Possible effects of the measure of false-vacuum eternal inflation are discussed below.} as initial states for subsequent tunneling through the moduli potential barrier, as they do not require uphill tunneling for their respective population.

We end up in the following situation. In all regions of the landscape where $b$ varies freely over moduli space between $b<1$ and $b>1$, the small-field and large-field regime contained as one-to-one in all large-field models of string theory are populated equally by tunneling. Therefore, the large-field mechanism by itself does not predict any bias towards small-field or large-field inflation, respectively.

The exception to this statement are possibly special regions in the landscape where microscopic constraints may impose $b < 1$ for, say, certain classes of compactifications. There is no small-field inflation at all in the large-field models in such a landscape region. However, the population of all possible large-field models in this region must proceed via uphill tunneling from $\phi_+=0$. We can then apply the same type of calculation as before to axion monodromy large-field models with varying monomial power, and suppressed instanton corrections. In these special landscape regions the above result implies a hierarchy among the large-field models in string theory. This hierarchy exponentially favors the models with the smallest monomial power $p$
\beq
\label{DOOM2}
\frac{P(\Delta\phi_{large-field,
\,\phi^p})}{P(\Delta\phi_{large-field,
\,\phi^{p'}})}\lesssim \frac{e^{-\frac{\Delta\phi_{large-field,
\,\phi^p}^2}{M_{\rm P}^2}}}{e^{-\frac{\Delta\phi_{large-field,
\,\phi^{p'}}^2}{M_{\rm P}^2}}}\sim e^{-2 N_e (p-p')}\ll 1\quad{\rm for}\quad p>p'\quad.
\eeq

The question whether small-field or large-field inflation is dominant in the landscape therefore has no answer within the large-field mechanism of string theory itself. The situation is better only for those presumably small regions of the landscape, which are characterized by bounded instanton corrections such that among them the large-field mechanism does not contain a small-field regime. We are thus forced to look to the far wider class of small-field models outside the range of the large-field mechanism, and we need to count -- and eventually weight that count by the combined dynamics of tunneling and eternal inflation.

We may now ask about conceivable loopholes in the line of thought of this section. One known alternative to extending the field range seen by the potential energy of an otherwise periodic single axion, by definition, via monodromy, is an assistance effect of many sub-Planckian axions, known as 'N-flation'~\cite{Dimopoulos:2005ac}. For this proposal to succeed, however, all of the axions used need a shift symmetry of similar quality as in single-field monodromy. The minimum of the combined effective multi-axion potential thus is similarly decoupled from vacuum-hopping in the moduli potential as in the single-field case -- again requiring super-Planckian uphill quantum diffusion in the axion valleys to enable entering the inflationary multi-axion valley via tunneling. This results in the same exponential bias against N-flation large-field models, as for single field axion monodromy setups.

Next, it is clear that the minimum of the inflaton potential in large-field models is almost decoupled from the moduli potential by virtue of the shift symmetry, but not completely so. Assume therefore that the minimum of the inflaton potential shifts by $\delta\phi\ll M_{\rm P}$ each time a next-neighbour vacuum transition executes in the moduli potential. Then there will possibly be multi-tunneling paths through the landscape which transport us into a progenitor false-vacuum axion valley with a minimum of the axion potential $\phi_{min.}\simeq \phi_{60}$ compared to the minimum of the final 'our-world' axion valley with $\phi_{min.}=0$. A tunneling jump from such a progenitor valley will not require $\phi$ quantum diffusing up the hill by a super-Planckian field range, and would thus cause no relative suppression compared to the instanton-tuned small-field sub-setup. However, as the shift of the axion minimum $\delta\phi\ll M_{\rm P}$ is very small, we will need many such tunneling jumps to get us into the right progenitor false-vacuum axion valley. As the maximum total potential difference $\Delta V_{tot}$ crossed by $N$ such jumps prior to arriving in the right progenitor valley in the controlled region of the landscape is $\Delta V_{tot.}<M_{\rm P}^4$ we have an average potential energy difference per jump of 
\beq
\Delta V^{(N)}\sim \frac{\Delta V_{tot.}}{N}\sim \frac{1}{N}\,M_{\rm P}^4\quad.
\eeq
In the limit of large $N$ we get $\Delta V_N\to 0$. Then the thin-wall limits holds for the $N$ successive CDL bounces which implies
\beq
S^{(N)}_E\xrightarrow[N\to\infty]{}\infty\quad\Rightarrow\quad \Gamma^{(N)}\sim e^{-S_E^{(N)}}\xrightarrow[N\to\infty]{}0\quad.
\eeq
As the full amplitude will be
\beq
\Gamma_{N\,jumps}\sim (\Gamma^{(N)})^N\sim e^{-N\,S_E^{(N)}}
\eeq
this will be exponentially suppressed for a super-Planckian $\Delta\phi_{60}\gg M_{\rm P}$ compared to a small-field model requiring $\Delta\phi_{60}\lesssim M_{\rm P}$, as then $N_{large-field}\gg N_{small-field}\gg 1$. This last hierarchy holds because the shift of the axion minimum due to a single-jump in the moduli potential is $\delta\phi\ll M_{\rm P}$. We have also assumed here that each of the $N$ jumps traverses on average the same typically sub-Planckian field interval in moduli space, with the interval length not or only weakly depending on $N$. This seems to be reasonable as long as the number of jumps $N$ is small compared to number $N_{vac.}\gtrsim 10^{500}$ the landscape must possess in order to allow for a weakly-anthropic explanation of the present-day small vacuum energy.

\newpage

\section{Counting ...}\label{sec:counting}
We have seen in the last section that within the largest part of the landscape the mechanism for large-field inflation in string theory provides for large-field and small-field inflation regimes with comparable likelihoods because they get populated evenly by CDL tunneling. We will see that such a democracy between large-field and small-field inflation seems to hold generally across the landscape as far as the dynamics of eternal inflation and tunneling are concerned -- as long as we post-condition our analysis on the regions of parametrically small-c.c.~dS vacua and inflationary regions which are cosmologically viable and permit an anthropic explanation of the observed extremely small c.c.~along Weinberg's argument. We start with discussing the effects of tunneling and eternal inflation together with anthropic post-conditioning, and then proceed to discuss attempts at counting the different inflationary realizations on the landscape.


\subsection{Eternity}\label{sec:eternal}

There a few salient facts concerning the dynamics of tunneling and false vacuum eternal inflation which seem to hold across the whole landscape:

\begin{itemize}



\item\label{itemI} I) All inflationary regions in the landscape will eventually get seeded via tunneling. The meta-stable dS vacua will undergo false-vacuum eternal inflation.

\item\label{itemII} II) The analysis of the dynamics of tunneling and eternal inflation in the landscape must be conditioned to those regions where Weinberg's anthropic explanation of today's extremely small positive cosmological constant (c.c.) is viable. Hence, we must confine ourselves to regions where an exponentially large number of vacua with small positive vacuum energy is efficiently populated.

\item\label{itemIII} III) The vast majority of the extrema in the scalar potential comes from the $p-$form flux discretuum. Hence, tunneling between vacua typically involves flux jumps. Transitions involving flux jumps of just a few units -- that is, transitions between vacua \emph{adjacent in flux space} - typically involve large differences in the vacuum energy.\footnote{As an example, take the Guka-Vafa-Witten flux superpotential $W=\int_{\cal M}G_{(3)}\wedge\Omega$ on  a warped Calabi-Yau ${\cal M}$. A tunneling transition involving, say, the change of a single flux quantum then typically implies a $|\Delta W|={\cal O}(1)$, and thus a large $|\Delta V|$ of order of the KK scale (see e.g.~\cite{Bousso:2000xa}).}

\item\label{itemIV} IV) The scalar potential on the landscape is a function on a high dimensional field space (we typically have $\#(moduli)={\cal O}(100\ldots 1000)$).

\end{itemize}

These will determine the ensuing sketch of an argument for inflationary democracy on the landscape. Let us begin with the properties I and II. These together rule out population of cosmologically viable small-c.c.~vacua by up-hill tunneling from an extremely small c.c.~dS vacuum, as such up transitions are punished by an an exponentially suppressed transition rate 
\beq
\Gamma_{up\;transition}\sim e^{-24\pi^2/V_{smallest\;dS}}\quad.
\eeq
This situation could only reverse itself if large-c.c.~dS vacua are more than exponentially rare, or down-hill tunneling from large-c.c.~dS vacua were forbidden.

Property III prevents this from happening. Furthermore,  the characteristics III and IV combined tell us that the immediate field space neighbourhood of any small-c.c.~dS vacuum will almost everywhere consist of large potential barriers and adjacent large-c.c.~dS vacua. Therefore, population of the small-c.c.~dS vacua will happen almost everywhere by direct and (compared to up transitions from extremely small-c.c.~vacua) fast down-hill tunneling transitions from adjacent large-c.c.~dS vacua. Combined with I and II we also see, that such fast down-hill tunneling is necessary in order to have a shot at efficiently populating an exponentially large number of small-c.c.~dS vacua for Weinberg's argument to work.

We can now compare this with the dynamics of eternal population. Appendix~\ref{sec:AppB} concerns itself exclusively with the one relevant aspect here -- the typical vacuum energy of the progenitor dS vacua. The discussion of this aspect follows in particular~\cite{Linde:2006nw}. The upshot can be summarized as this: For global measures with full volume weighting the progenitor is the \emph{largest-c.c.} metastable dS vacua. This is the result of the exponential 3-volume reward driving the progenitor Hubble parameter to be as large as possible. All other measures free of obvious paradoxa (global measures without exponential volume reward, such as the scale factor measure, or local measures, such as the causal diamond measure) will see the the landscape populated from the \emph{longest-lived} progenitor dS vacuum.

From the discussion above, and in Appendix~\ref{sec:App} we know that parametrical longevity of dS vacua is achieved once it has no down-hill tunneling paths accessible, and an exit thus will proceed by an up transition. Therefore, the longest-lived progenitor of a given region in the landscape will be a dS vacuum of somewhat small c.c.~which can exit only via up-hill tunneling to another dS vacuum of near-Planckian c.c.~Note that typically the c.c.~of the progenitor will be just somewhat small compared to the Planck (or string) scale. In most cases it will not have the extremely small values relevant for cosmologically viable vacua because the up transition rate behaves as
\beq
\Gamma_{longest-lived;progenitor,\;up}\sim e^{-\frac{24\pi^{2}}{V_{longest-lived;progenitor}}}\quad.
\eeq
Note, that the remaining extreme cases where the longest-lived progenitor has extremely small-scale c.c.~and thus does not need a near-Planckian intermediate state to guarantee longevity of the up transition, are taken care of by the requirement of Weinberg's argument to work. Namely, those extremely small-c.c.~progenitors which do not exit via up-hill tunneling into very high-scale dS vacua, will not efficiently populate an exponentially large number of very small-c.c.~descendant dS vacua, and thus are ruled out anthropically.

We therefore expect in all relevant classes of inflationary measures that the population of an exponentially large number of cosmologically viable descendant vacua in the landscape involves as their immediate predecessor a meta-stable dS of very high-scale c.c.~




\subsection{Democracy -- tumbling down the rabbit hole}\label{sec:democracy}

We know at this point that all cosmologically and anthropically viable small-c.c.~regions of the landscape have a very high-scale c.c.~dS vacuum as their immediate progenitor. If we can establish in addition that the down-hill tunneling rate from a high-scale c.c.~vacuum does not depend on the vacuum energy of a small-c.c.~target dS vacuum, then we can show that tumbling down from the anthropically selected high-scale c.c.~immediate progenitor into the small-c.c.~regions of the landscape provides a flat prior for those vacua. Tunneling into the small-c.c.~part of the dS landscape from the anthropically selected high-scale progenitor proceeds democratically.

The independence of the Euclidean bounce action $B=S_{E}(\phi)-S_{E}(\phi_+)$ for CdL tunneling from small changes of the vacuum energy $V_-$ of the target small-c.c.~dS vacuum can be demonstrated for a general potential as well as for the thin-wall limit. The outcome of the preceding discussion dictates a hierarchy of vacuum energies $V_T>V_+\gg V_-\geq 0$. $V_T$ denotes the height of the potential barrier which separated the progenitor dS vacuum with large c.c.~$V_+$ from one of the cosmologically viable descendant dS vacua with very small c.c.~$V_-$. According to~\cite{Kachru:2003aw} we can write
\beq
S_E(\phi)=-2\pi^2\,\int d\xi\,\rho^3(\xi)\,V(\phi)\quad.
\eeq
We tunnel into different descendant small c.c.~dS vacua whose vacuum energies $V_-,V_-'$ differ by a still parametrically small amount $|\Delta V_-|\ll V_+$. Hence, we immediately get that the negative term in $B$
\beq
S_E(\phi_+)=-\frac{24\pi^2}{V_+}\cdot\left[1+{\cal O}\left(\frac{\Delta V_-}{V_+}\right)\right]
\eeq
gets only a negligible correction from the variation of the vacuum energy among the many small-c.c.~descendant vacua. The Euclidean action for the bounce solution itself we will approximate from above, and find the same parametrical result. We have roughly
\beq
S_E(\phi(\xi))\sim \Delta\xi\,\rho_{max}^3\,\big(V_T+{\cal O}(\Delta V_-)\big)
\eeq
where $\Delta\xi$ is the Euclidean time elapsed while $\rho(\xi)$ increases from zero to $\rho_{max}$ and decreases back to zero. From the CdL Einstein equation we get $\rho_{max}$ at $d\rho/d\xi=0$ to be
\beq
\rho_{max}\lesssim \sqrt{\frac{3}{V_+}}\quad.
\eeq
Moreover, we have $\Delta\xi\lesssim H_+^{-1}=\sqrt{3/V_+}$. Hence, we arrive at
\beq
S_E(\phi)\sim-\frac{1}{V_+}\cdot\left[1+{\cal O}\left(\frac{\Delta V_-}{V_+}\right)\right]
\eeq
with the same parametrically suppressed correction. One can further show using the field e.o.m. that also the bounce solution $\phi(\xi)$ gets parametrically small corrections of ${\cal O}(\Delta V_-/V_+)$.

In the thin-wall limit where we have not just $V_T>V_+\gg V_-\geq 0$ but $V_T\gg V_+\gg V_-\geq 0$, we can calculate the influence of a shift $|\Delta V_-|\ll V_+$ in $V_-$ on $B$ exactly and get
\beq
B=\frac{27\pi^2 T^4}{2 V_+^3}\cdot\frac{1}{(1+3T^2/4V_+)^2}\cdot \left[1+{\cal O}\left(\frac{\Delta V_-}{V_+}\right)\right]
\eeq
in agreement with the general result above.

Summing up, we know that the anthropically efficient population of a cosmologically viable small-c.c.~region of the landscape proceeds via tunneling from a very high-scale c.c.~dS vacuum, and the down-hill tunneling rate into such small-c.c.~vacua is independent from the varying vacuum energy of the many small-c.c.~vacua. This implies that the tunneling rates get their variation from the distribution of barrier shapes -- with the two main parameters height $V_T$ and thickness $\Delta\phi$ -- across the many down-hill tunneling paths from the high-scale progenitor to the exponentially many small-scale c.c.~descendants.

Since the rates are \emph{independent} from the varying c.c.~of the low-scale dS vacua, we can thus \emph{average} over the barrier shapes. Effectively, for small-c.c.~regions satisfying anthropically efficient population the neighbourhood of each small-c.c.~dS vacuum will show the same statistical distribution of high-scale progenitors and barrier shapes which allows for averaging over them. We can thus write the down-hill tunneling rate for small-c.c.~regions satisfying anthropically efficient population as
\beq\label{eq:finalfeeding}
\Gamma_{large-c.c.\to small-c.c.}\sim e^{-\langle\Gamma(V_+)\rangle_{barrier\,shapes}}
\eeq
where the barrier shape average $\langle\Gamma(V_+)\rangle_{barrier\,shapes}$ will be a function of the immediate progenitor's vacuum energy $V_+$ only. Again, the immediate progenitor denotes either the highest dS minimum for fully volume-weighted global measures, or the intermediary exit vacuum of the longest-lived dS vacuum for all other measures.

Eq.~\eqref{eq:finalfeeding} describes a central result which we expect to hold across all of the landscape: Namely, to be efficient enough for Weinberg's argument to work, the population of cosmologically viable small-c.c.~regions of the landscape 
proceeds via down-hill tunneling from very high-scale c.c.~progenitors, and this process populates the small-c.c.~vacua democratically, placing no prior due to tunneling. 

Let us compare this to the discussion in section~\ref{sec:argument}. From there we know that the small-field regime inside stringy large-field models has to use the lowest lying instanton-induced inflection point, because doing otherwise would get the inflaton trapped in local minima at lower-lying inflection points. However, recently the absence of the overshoot problem was shown for tunneling-born small-field models~\cite{Dutta:2011fe}. This implies that we can tunnel into the small-field regime of a large-field model as far out and high (in potential energy) above the inflection point as we wish to. We can therefore, using the notation of section~\ref{sec:argument}, seed both a small-field regime within a large-field model of string theory, as well as its own large-field regime by tunneling from an instanton-induced local minimum close to $\phi_+\simeq \Delta\phi_{large-field}\gg M_{\rm P}$. 

\begin{figure}[t!]
\begin{center}
\includegraphics[width=0.52\textwidth]{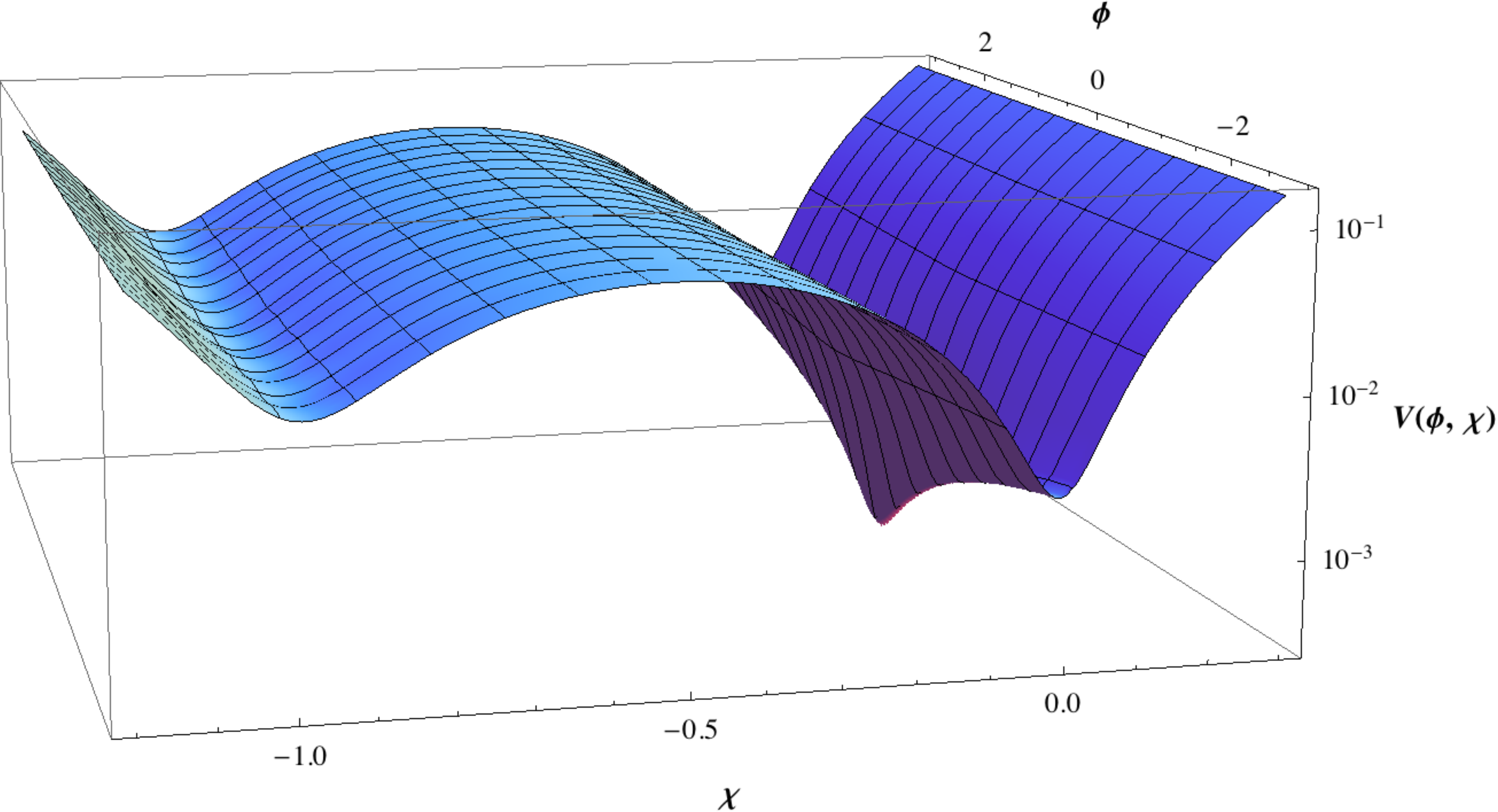}
\includegraphics[width=0.46\textwidth]{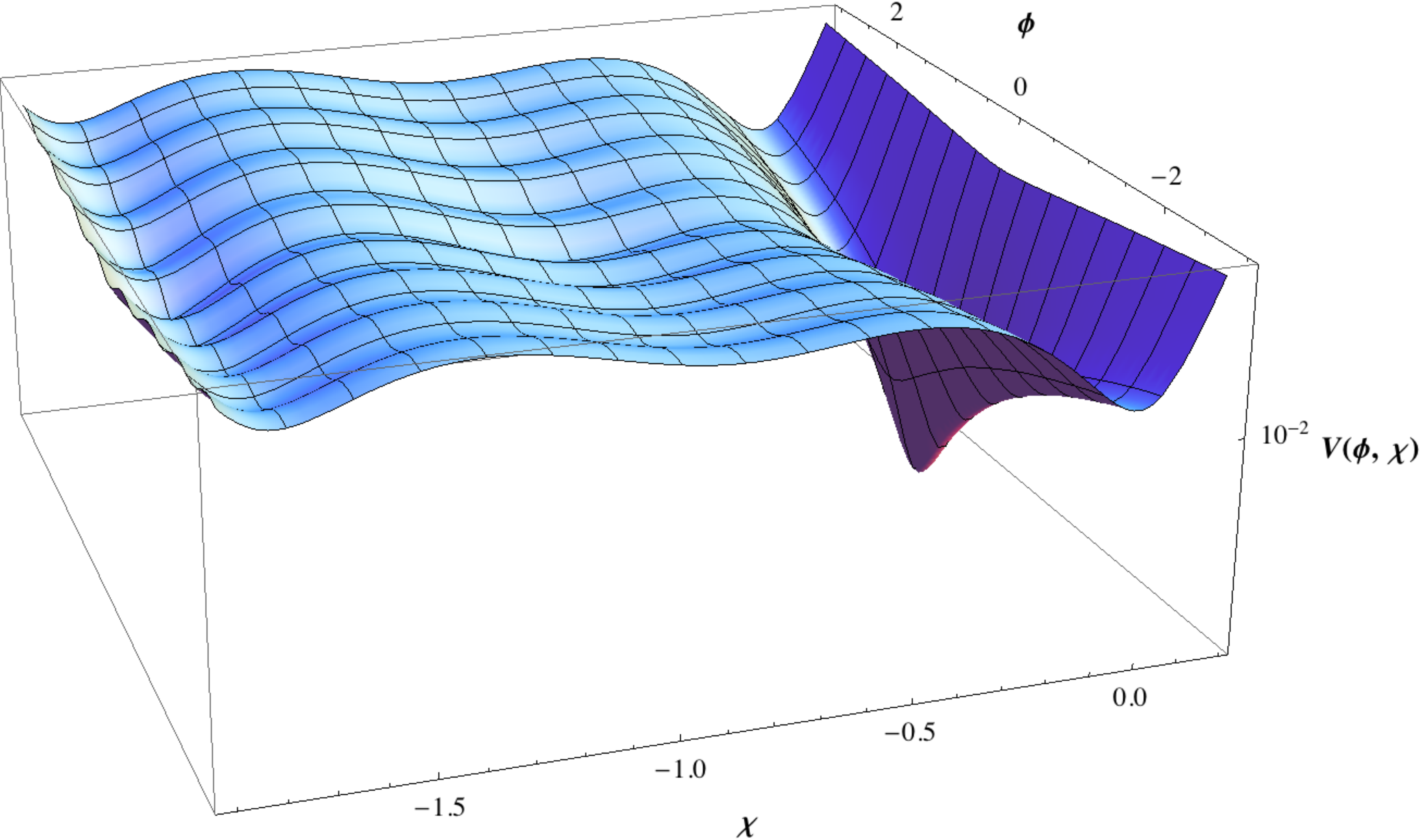}
\end{center}
\refstepcounter{figure}\label{Vplot4}
\vspace*{-.2cm} {\bf Figure~\ref{Vplot4}:} A generic large field potential $V(\phi,\chi)$ in logarithmic scale for
the canonically normalized inflaton field $\phi$ and modulus $\chi$ (in Planck units). The magnitude of the instanton corrections $b\simeq 1$ typically varies over moduli space. {\bf Left:} The schematic case for a global volume-weighted measure of eternal inflation is shown. The progenitor is given by the largest-c.c.~meta-stable dS vacuum. Due to the axionic shift symmetry the largest-c.c.~progenitor extends along a valley broken up into a series of shallow equidistant minima by instanton corrections. As these are exponentially volume suppressed, and the largest-c.c.~dS vacuum typically will be almost Planckian in energy, the instanton ripples are very shallow. Evidently, this produces a series of equidistantly spaced highest-c.c.~progenitor dS vacua which all have the down-hill tunneling rate into the slow-roll valley at $\chi=0$, and are thus equivalent. {\bf Right:} The schematic case for a local measure of eternal inflation is shown. The left-most valley of vacua forms the longest-lived progenitor vacua. The instanton-induced ripples are depicted larger here, as the longest-lived progenitor typically have high-scale c.c.~but are not necessarily almost Planckian. The progenitor vacua exit by passing via up-hill tunneling through the metastable mediator valley in the middle. We see that largely equivalent parallel paths connect the equidistantly space local minima which are produced by the instanton effects in the direction of the axion in both the progenitor and the mediator valley. Down-hill tunneling again proceeds from the mediator valley with equal rates for all the local minima in the axion direction as the rate is independent of the tiny vacuum energy of the descendant valley at the right or its tiny vacuum energy variation.
\end{figure}

Now take into account the anthropically required democracy in down-hill tunneling which feeds our slow-roll inflationary regions in a cosmologically viable region of the landscape. We are in the regime of a flat tunneling prior, because cosmological viability forces all slow-roll inflationary vacuum energies by COBE normalization to have $V\lesssim 10^{-10}$ which is already in the small-c.c.~regime compared to the progenitor vacua.
Hence, down-hill tunneling from the high-scale progenitors will populate all false minima in the false vacuum valley, and also all starting points in the true vacuum valley  $\phi_+$ evenly. As this includes values $\phi_+\simeq \Delta\phi_{large-field}\gg M_{\rm P}$ , and we can seed both the small-field and the large-field regime from the \emph{same} $\phi_+\simeq \Delta\phi_{large-field}\gg M_{\rm P}$, this will populate both the small-field and the large-field regime of every string theory large-field model equally. At the same time, the same argument leads to equal population of all other small-field saddle points outside the large-field mechanism class. Hence, we conclude that the dynamics of eternal inflation and vacuum tunneling transitions realize both small-field and large-field inflation with a flat prior, when conditioned on cosmologically and anthropically viable descendant regions. Fig.~\ref{Vplot4} displays two examples which show schematically the similarity of the global and local measures, and the democracy in down-hill tunneling that ensues from requiring cosmologically and anthropically viable descendant vacua, and the vacuum structure the shift symmetry enforces on the progenitors in the axion direction.

As presented, this argument still has a possible loophole. Everything we said holds strictly true if the lowest-lying 1st inflection point is the only one suitable for small-field inflation. However, the latter statement is only valid for convex large-field  potentials $V_0(\phi)\sim \phi^p\;,\;p\geq 1$. For concave potentials $V_0(\phi)\sim \phi^p\;,\;0<p<1$, however, which are still large-field for $p\gtrsim 0.1$, you could equally well have small-field on the $n^{th}$ inflection point at $\phi_{n^{th}\,inflection\,point}>M_{\rm P}$. This follows because almost all (except for the first few ones at $\phi\ll M_{\rm P}$ where $V_0(\phi)\sim \phi^2$ again) of the lower-lying inflection points will have $b<1$ as $V_0(\phi)$ is concave. The number of inflection points generated by the instanton correction within $\Delta\phi_{60,large-field}$ is given by
\beq
N_{inflection\,point}\sim  \frac{\Delta\phi_{60,large-field}}{f}\quad.
\eeq
We have already discussed that $f\lesssim M_{\rm P}$ in string theory, and in concrete models of 5-brane axion monodromy one gets e.g.~\cite{Flauger:2009ab}
\beq
 \frac{g_s^{1/4}}{(2\pi)^{3/2}\sqrt{\cal V}}<\frac{f}{M_{\rm P}}<g_s\,\frac{\sqrt{3}}{2}\quad.
\eeq
Here ${\cal V}$ is the warped volume of the internal manifold in units of $\alpha'$. Thus, $N_{inflection\,point}\gg 1$ typically, and for concave models each of them is available for slow-roll tuning. This might lead us to conclude towards a counting bias towards small-field models. However, the same fact $N_{inflection\,point}\gg 1$ also implies that the inflection points are spaced densely compared to the evolution of $V_0(\phi)$. Therefore, if the $n^{th}\;,\;n\gg1$ inflection point is tuned flat ($b=1$), than its neighbours will have $b\simeq 1$ to very good degree, too. This leads to a wide field range over which the slow-roll parameters $\epsilon,\eta$ have sizable oscillations. They, in turn, imprint themselves as large oscillations on the 2-point function power spectrum of the curvature perturbation generated, which are severely bounded by the observed the CMB~\cite{Flauger:2009ab}. In particular, these limits imply $b\ll 1$~\cite{Flauger:2009ab}, and therefore small-field inflation starting from a high-lying $b=1$-tuned inflection point would give a universe with the wrong CMB. This removes all such small-field candidates except the ones starting from the few lowest-lying inflection points from the comparison with the large-field regime. So, demanding consistency with the observed CMB leads us back to the conclusion already drawn above.

\subsection{Vacuum energy distribution}\label{sec:vacdistro}

This democratic result is to be compared with the product of the number frequency distribution of the vacuum energy of inflationary regions in the landscape, and the number frequency distribution for different inflationary model classes on the landscape. We start here the vacuum energy distribution. This prior is relevant as the field range of a given inflationary region implies a posterior constraint on the admissible vacuum energy range from the COBE normalization of the CMB fluctuations. Otherwise one could average over all occurring vacuum energies because the mechanisms for realizing small-field or large-field inflation in string theory do not depend strongly on the potential energy scale realized. This poses a danger if the prior number frequency distribution of the vacuum energies were to scale like
\beq
P_{V_{infl.}}\sim e^{\frac{1}{V_{infl.}}}
\eeq
as this would offset the hierarchy introduced by CdL tunneling discussed before. We lack calculational access to large swaths of the landscape, so we can only look at estimates of number frequency distributions of vacuum energies in corners where we have access. In one such corner, flux compactifications of type IIB string theory on warped Calabi-Yau threefolds, space-time supersymmetry can be used to estimate the distribution of vacuum energies among supersymmetry breaking vacua where the fluxes stabilize the moduli. The relevant distribution computed there is the distribution of the supersymmetry breaking scale $M_S^2$~\cite{Ashok:2003gk,Susskind:2004uv,Douglas:2004qg,Denef:2004ze,Denef:2004cf} (for a review, see e.g.~\cite{Douglas:2006es}). We have $M_S^2\sim F$ in terms of the supersymmetry breaking F-terms, and the upper limit $V_{max}$ of the positive vacuum energy of a given vacuum is related to the F-terms as $V_{max}\sim F^2$. Hence, we can estimate the large-scale distribution of vacuum energies as the one given for $M_S^2$ provided that no strong tuning of the vacuum energy has been selected for (the situation relevant for inflation). According to the results of~\cite{Susskind:2004uv,Douglas:2004qg,Denef:2004cf}, this leads to
\beq
P_{V_{infl.}}\sim M_S^2\sim F\quad.
\eeq
In flux compactifications we expect $F\sim W$ to have a flat distribution. This implies a flat number density distribution
\beq
dP_{V_{infl.}}=const.
\eeq
for the vacuum energy.

\subsection{Multitude}\label{sec:manymany}

We are thus left with estimating the number frequency distributions of small-field and large-field inflation mechanisms in the string landscape. Let us start with the generic small-field models on the landscape (i.e. those which do not arise from the instanton contributions to large-fields models with axion monodromy). Most of these occur 'accidentally', that is, in vacua where the microscopic parameters such as fluxes, result in a local inflationary slow-roll flat dS saddle point of the moduli potential which can drive inflation. Barring further constraints, we can as a very rough approximation model the landscape (outside the symmetry-protected large-field mechanism occurrences) as a random potential for an $N$-dimensional scalar field space.\footnote{For recent work on modeling the string landscape this way, see e.g.~\cite{Aazami:2005jf,Frazer:2011tg,Agarwal:2011wm,Marsh:2011aa,Chen:2011ac}.} We then need to determine how many slow-roll flat dS saddle points we statistically expect in such a description. This question has been dealt with in a work by Aazami \& Easther~\cite{Aazami:2005jf}, where the propose to model the landscape as a random potential given by
\beq\label{randpot}
V(\chi_i)=\sum_{i=1}^N f_i(\chi_i)+\sum_{i\neq j}c_{ij} \chi_i\chi_j\quad.
\eeq
We then have two cases. To describe them, let us estimate the scale of the cross couplings as $c_{ij}\sim M^4/M_{\rm P}^2$.

At first, we can now look at the case where $M\ll M_{\rm P}$ of very small cross couplings. If each of the functions $f_i$ has $\alpha_i\geq 1$ extrema, then the total number of extrema due to the lack of cross-terms is given by
\beq
N_{extr.}=\Pi_{i=1}^N \alpha_i=\alpha^N
\eeq
where $\alpha$ denotes the geometric mean of the $\alpha_i$. With $N$ easily being of ${\cal O}(10^3)$ in the moduli space of string theory, even an $\alpha$ as close to unity as, say, $1.1$ would imply ${\cal O}(10^{100}\ldots10^{1000})$ critical points in the landscape. Saddles among these critical points are classified by a Hessian which does not have all positive eigenvalues. The theory of random matrices then tells us that for an $N\times N$ symmetric almost diagonal matrix, each choice and permutation of eigenvalue signs occurs statistically with a frequency approaching $1/2^N$~\cite{Aazami:2005jf}. A local minimum, represents just a single choice among all possible choices and permutations of eigenvalue signs.  Thus, the number of saddle points (including local maxima) in our model landscape is
\beq
N_{saddle}=\alpha^N\,\left(1-\frac{1}{2^N}\right)\simeq\alpha^N
\eeq
while we get only
\beq
N_{min.}=\left(\frac{\alpha}{2}\right)^N\quad.
\eeq
Almost all of the critical points are saddles. For inflationary purposes we may wish to restrict our attention to class of saddles with just \emph{one} negative eigenvalue, as these guarantee single-small-field inflation. For small cross couplings their number is
\beq
N_{single-field\, saddle}=N\,\left(\frac{\alpha}{2}\right)^N\quad.
\eeq
There is by now ample evidence  that the landscape contains, even in the small calculable sectors, more than ${\cal O}(10^{100})$ local minima. This tells us that we have to put $\alpha={\cal O}(4)$, and thus there are easily more than ${\cal O}(10^{100})$ single-field saddle points available.

Keeping this estimate for $\alpha$, the number of local minima and single-field saddle points begins to decrease super-exponentially compared to the above results only in the extreme opposite case where $M\sim M_{\rm P}$ (i.e. when the cross couplings are of the same order as the $f_i$ themselves). In this case the Hessian of the extrema of $V$ becomes a general symmetric matrix. If the coefficients in the $f_i$ and the $c_{ij}$ are drawn from a normal distribution, then the Hessian of the extrema of $V$ is a symmetric matrix drawn from Gaussian Orthogonal Ensemble. Its eigenvalue distribution obeys the Wigner semi-circle law, i.e the eigenvalue density $E(\lambda)$ is
\beq
E(\lambda)=\left\{\begin{array}{c}\frac1\pi \sqrt{2N-\lambda^2} \\0\quad{\rm for}\;\;\lambda>\sqrt{2N}\end{array}\right.\quad.
\eeq
One can then show~\cite{Aazami:2005jf} (see also more recently~\cite{Marsh:2011aa,Chen:2011ac}) that the probability to have a local minimum (i.e. all positive eigenvalues of the Hessian) or a single-field saddle point is given by
\beq
P_{min./single-field\,saddle}\sim e^{-\frac{N^2}{4}}\quad.
\eeq
As shown in~\cite{Aazami:2005jf}, already a separation of scales between the $f_i$ and the cross couplings as small  as 2 orders of magnitude is enough to sit safely within the first case discussed above, giving exponentially many local minima and single-field saddles potentially suitable for inflation.

In general, one expects these two cases to appear mixed together in that a few cross couplings may appear with $M\gtrsim 10^{-2} M_{\rm P}$ while most of them will be at smaller scales. The corresponding Hessian will then be approximately band diagonal, but the count of single-field saddle points will remain exponentially large of ${\cal O}(10^N)$, because band width is generically expected to be small compared to $N$.

One may condition this analysis on subsectors of the landscape which potentially allow for low-energy space-time supersymmetry. In such a situation the random potential over moduli space should be replaced by a random supergravity, i.e. random choices for the Kahler and superpotential of the moduli. Moreover, in such a supersymmetric sector of the landscape we should envision for a large number $N_H<N$ of moduli being stabilized supersymmetrically at a large mass scale (flux stabilization of complex structure moduli and the axio-dilaton in type IIB on a warped Calabi-Yau provides a large class of examples), while supersymmetry breaking occurs together with the stabilization of the small number $N_L$ of remaining moduli at a parametrically smaller mass scale. Both effects have been studied in detail in~\cite{Marsh:2011aa} with the result that the probability to have a local minimum (i.e. all positive eigenvalues of the Hessian) is given by
\beq
P_{min.}\sim e^{-c_L\,N_L^p}\quad.
\eeq
Here $1<p<2$ and $c_L$ is an ${\cal O}(1)$ number which can be estimated with random matrix methods~\cite{Marsh:2011aa}. The total number of local minima (i.e. all positive eigenvalues of the Hessian) in a given sector with $N>N_H\gg N_L$ moduli then remains still exponentially large
\beq
N_{min.}\sim e^{c_H N_H}e^{-c_L\,N_L^p}\gg1\quad.
\eeq

The upshot is that we will get in a landscape with $N$ scalar degrees of freedom typically ${\cal O}(10^N)$ meta-stable dS minima. The fraction $\beta_{saddle}$ of them which constitute single-field saddle points potentially suitable for inflation we do not know so far. From the existing random matrix studies~\cite{Marsh:2011aa} so far it is not clear whether there will be more or less single-field saddle points than meta-stable dS minima. What we do know about is the cost of flatness of such a saddle point. The fraction of them which are locally flat enough to support $60+$ e-folds of slow-roll inflation will be determined by the fraction of volume in microscopic parameter space, such as fluxes, which yields sufficiently flat saddle points. According to work done by~\cite{BlancoPillado:2004ns,Baumann:2007ah,Linde:2007jn,Agarwal:2011wm} this imposes a fine-tuning cost of typically ${\cal O}(10^{-8}\ldots10^{-2})$ in the space of single-field saddle points, with~\cite{Agarwal:2011wm} most recently finding this suppression for warped D3-brane inflation to be of ${\cal O}(10^{-5}\ldots10^{-3})$. This cost is negligible compared to the quasi double-exponential number ${\cal O}(10^N)$ of dS minima, as $N={\cal O}(100\ldots 1000)$.

The next step consists of counting the realizations of the large-field mechanism in the landscape. The crucial differences to the small-field count above reside in the need for a shift symmetry and the functional 'fine-tune' characteristic for large-field inflation. The combination of both requires a realization of the large-field mechanism to have a distinctly projected-in axion field, with a distinctly chosen 'discrete' source of potential energy with non-trivial axion monodromy. This can work only for each given axion direction once-a-time, and can not, by definition, yield multiple locally-flat regions in each axion field space direction, because the potential energy source has to have monodromy and thus is of a fixed large-field functional form.

However, this leads to a crucial difference in counting. Projecting in a suitable RR-form axion field, and supplying it a source of potential energy with axion monodromy, constitute discrete choices selecting a whole manifold for compactification. On each such manifold there is still a potentially large discretuum of vacua generated by the available choice of fluxes used in moduli stabilization. If the number of moduli is large the available flux discretuum will be only insignificantly changed by imposing the condition of e.g. projecting in a suitable axion. Therefore, a large fraction of all available flux dS vacua \emph{on a given} manifold of compactification will lead to a axion monodromy large-field inflation~\footnote{I.e., they are available as viable exits of inflation, and they remain stable in presence of the large-field inflationary vacuum energy.}, if the manifold itself was chosen correctly, while on the same manifold only a certain fraction of all available flux dS vacua will constitute a sufficiently tuned small-field inflationary saddle point. We do not yet know whether the latter are more abundant than dS minima or not.

The estimation of the number frequency distributions of generic small-field saddle points and axion monodromy large-field regions requires us therefore to determine the abundance of small-field single-field saddle points relative to the one of the dS minima, sum over all compactification manifolds, and determine the fraction of them which allow for projecting in a suitable axion and supplying it potential energy with axion monodromy. In general, we do not have (yet) sufficient calculational access into the landscape to do so.

\subsection{An accessible sector of landscape}\label{sec:CYlandscape}

Still, it is potentially possible to answer a more modest form of the same question for a known and calculable sector of the landscape. One such example is the landscape of flux vacua on warped Calabi-Yau (CY) orientifold compactifications of type IIB string theory. This sector of the landscape may be of additional interest, as any possible strong number frequency bias arising there would tie the result to a possible detection of low-energy supersymmetry by virture of being most naturally realized in Calabi-Yau compactifications.

On this sector of warped fluxed CY compactifications of type IIB we can now specify the parameters entering the number frequency distributions of inflationary mechanisms a bit more precisely. In particular, we have $N_H\geq h^{2,1}+1$ and $N_L\leq h^{1,1}_+$, where $h^{2,1}$ denotes the number of complex structure moduli on a given CY 3-fold stabilized supersymmetrically at a high mass scale by fluxes together with the axio-dilaton, while $h^{1,1}_+$ counts the number of Kahler moduli. From the last section we have on each CY an estimate for number of all critical points of the moduli potential
\beq
N_{i,cr.}\sim e^{c_{h^{2,1}_i} h^{2,1}_i}
\eeq
while the fraction of meta-stable minima is
\beq
\beta_{i,dS-vac.}\sim e^{-c_{h^{1,1}_{i,+}}\,\left(h^{1,1}_{i,+}\right)^p}\quad.
\eeq
In terms of these we can now estimate the number of local minima $N_{i,\,min.}$ on a given CY 3-fold $i$
\beq\label{dSmincount}
N_{i,\,min.}\sim  N_{i,cr.}\cdot\beta_{i,dS-vac.}\gg1
\eeq
and the fraction of those which are sufficiently fine-tuned inflationary single-field saddle points
\beq\label{smallfieldcount}
N_{i,\,single-field\,saddle}\sim N_{i,cr.}\cdot\beta_{i,dS-vac.}\cdot\beta_{i,flat\;saddle}\cdot\left(1-\beta_{i,V^{\frac14}>10^{16}{\rm GeV}}\right) \quad.
\eeq
Here $\beta_{i,flat\;saddle}$ denotes the ratio of the number of inflationary flat single-field saddle points to the number of meta-stable dS minima. Note that we do not know a priori whether $\beta_{i,flat\;saddle}<1$ or $\beta_{i,flat\;saddle}>1$. A more detailed study of random matrix models along the lines of~\cite{Marsh:2011aa} may yield an answer to this question. 
The quantity of $1-\beta_{i,V^{\frac14}>10^{16}{\rm GeV}}$ denotes the fraction of such inflationary single-field saddle points with an energy scale small enough to support observationally viable inflation on a sub-Planckian field range. Finally, we now have to sum this over all CY 3-folds.  If we denote averages of the above quantities over the number of CY manifolds by dropping the label $i$, then we get
\beq\label{smallfieldcountTot}
N_{\,single-field\,saddle}\sim N_{CY}\cdot N_{cr.}\cdot\beta_{dS-vac.}\cdot\beta_{flat\;saddle}\cdot\left(1-\beta_{V^{\frac14}>10^{16}{\rm GeV}}\right) \quad.
\eeq

To determine the fraction of manifolds with axion monodromy inflation we have to multiply each term eq.~\eqref{dSmincount} with a factor $\delta$ which is either zero or unity depending on whether the given CY 3-fold has a projected-in RR $C_2$-form axion (or equivalently, $h^{1,1}_{i,-}=1$) and suitable source of potential energy with axion monodromy, everything placed inside a warped throat etc. For a conservative estimate we may ask to bound the number of CY's supporting axion monodromy by counting all those with $h^{1,1}_{i,-}\geq1$ as the most basic requirement ($\delta=\delta_{h^{1,1}_{i,-}\geq1}\in{0,1}$). We still do not know how to do this for all CY 3-folds. But, we may be able to do this for a large set (several million CY 3-folds) of examples given by their corresponding F-theory compactifications on an elliptically fibered CY 4-fold which are given as hypersurfaces in ambient toric spaces. This class of fluxed warped CY compactifications of type IIB is specified completely in terms of the discrete data of the GLSM description of the ambient toric spaces and hypersurfaces therein together with 4-form flux data. The discrete GLSM data then allows for determining for each choice whether $h^{1,1}_{i,-}\geq1$, and thus to determine the fraction $\beta_{h^{1,1}_{-}\geq1}$ of all CY's within this sample which support the basic requirement of large-field inflation. Next, we denote with $\langle h^{1,1}_-\rangle$ the average number of RR 2-form axions projected in on the elliptically fibered toric ensemble. Moreover, we have to restrict to the fraction $\beta_{i,V^{\frac14}>10^{16}{\rm GeV}}$ of axion monodromy realizations with sufficiently large energy scale to drive to correct amount of curvature perturbations. Hence, we write $N_{large-field}\leq N_{h^{1,1}_{-}\geq1}^{toric\,F-theory\,CY_4's}$ and
\beq
\left.\frac{N_{large-field}}{N_{single-field\,saddle}}\right|_{toric\,F-theory\,CY_4's}\leq \frac{N_{h^{1,1}_{-}\geq1}^{toric\,F-theory\,CY_4's}}{N_{single-field\,saddle}}\quad.
\eeq
Plugging in we thus get what we may call the  'landscape Drake equation'~\cite{DRAKE}
\bea\label{eq:finalresult}
\frac{N_{h^{1,1}_{-}\geq1}^{toric\,F-theory\,CY_4's}}{N_{single-field\,saddle}}&\sim& \frac{\sum_{i}N_{i,cr.}\cdot\beta_{i,dS-vac.}\cdot h^{1,1}_{i,-}\cdot \beta_{i,V^{\frac14}>10^{16}{\rm GeV}}\cdot\delta_{h^{1,1}_{i,-}\geq1}}{\sum_{i}N_{i,cr.}\cdot\beta_{i,dS-vac.}\cdot\beta_{i,flat\;saddle}\cdot\left(1-\beta_{i,V^{\frac14}>10^{16}{\rm GeV}}\right)}\nonumber\\
&&\nonumber\\
&=& \frac{\beta_{h^{1,1}_{-}\geq1}\langle h^{1,1}_-\rangle\beta_{V^{\frac14}>10^{16}{\rm GeV}}\sum_{i}N_{i,cr.}\cdot\beta_{i,dS-vac.}}{\beta_{flat\;saddle}\cdot\left(1-\beta_{V^{\frac14}>10^{16}{\rm GeV}}\right)\sum_{i}N_{i,cr.}\beta_{i,dS-vac.}}\nonumber\\
&&\nonumber\\
&&\nonumber\\
&=&\left.\frac{\beta_{h^{1,1}_{-}\geq1}\cdot\langle h^{1,1}_-\rangle\cdot\beta_{V^{\frac14}>10^{16}{\rm GeV}}}{\beta_{flat\;saddle}\cdot\left(1-\beta_{V^{\frac14}>10^{16}{\rm GeV}}\right)}\right|_{toric\,F-theory\,CY_4's}\quad.
\eea
The sums in these expressions run over the set of CY's denoted by $toric\,F-theory\,CY_4's$. This result assumes a flat number frequency distribution of vacuum energy. Arguments for this flat prior to arise in the context of type IIB flux compactification were reviewed above in section~\ref{sec:vacdistro}.\footnote{See, however, recent work on type IIA perturbative moduli stabilization~\cite{Chen:2011ac}, and type IIB Kahler uplifting~\cite{Sumitomo:2012wa}, where a product probability distribution over several individually flat microscopic variables seems to give a vacuum energy distribution peaked around zero. Such a non-flat would have to factored into the result above eq.~\eqref{eq:finalresult}, giving a bias towards small-field models with their smaller vacuum energies.}

Note that we do not know a priori whether $\beta_{flat\;saddle}<1$ or $\beta_{flat\;saddle}>1$. An estimate of the axionic in-projection cost $\beta_{h^{1,1}_{-}\geq1}$ seems feasible for the large sample of CY 3-folds described in F-theory as elliptic 4-folds given in terms of their GLSM data. It is conceivable that a study of random matrix models along the lines of~\cite{Marsh:2011aa} may yield in fact $\beta_{flat\;saddle}>1$. Then in virtue of $\beta_{h^{1,1}_{-}\geq1}$ bounding the number frequency of axion monodromy inflation from above, this would tell us to expect a negligible tensor fraction $r$ in the type IIB CY landscape to the extent that the occurrence of an observable tensor-to-scalar ratio $r\gtrsim 0.01$ is tied to the inflationary scale and thus to the existence and realization of large-field models of inflation.\footnote{The link between $r\gtrsim 0.01$ and large-field inflation is not watertight. On the one hand axion inflation can lead to additional highly non-Gaussian scalar perturbations sourced through the axion-photon coupling, which effectively suppresses $r$ even for large-field models~\cite{Barnaby:2010vf,Sorbo:2011rz}. Next, small-field inflation models can be (severely!) fine-tuned to produce $r\gtrsim 0.01$~\cite{BenDayan:2009kv,Hotchkiss:2011gz}. And finally, a small-field inflaton can source additional scale-invariant B-mode power through couplings to degrees of freedom (particles or strings) which get light at points of enhanced symmetry~\cite{Senatore:2011sp}, similar to the trapping mechanism~\cite{Kofman:2004yc,Green:2009ds}.}

 


\section{Discussion}
\label{sec:discuss}

Let us stop here to summarize again the crucial aspects of the story just told. Firstly, the premises laid out in subsections~\ref{sec:symm}, and~\ref{sec:scalars} together imply that the properties of the scalar field from string compactification require large-field inflation in string theory to take the form of axion monodromy. The axionic shift symmetry, rooted in the $p$-form gauge symmetry on the worldsheet, decouples the position of the minimum of the axion monodromy inflaton potential from the moduli potential. If this were otherwise, it would imply sizable non-gravitational couplings between the inflaton axion and the moduli which would invalidate the shift symmetry in the first place. Therefore, the many local minima of the moduli potential landscape share (almost) the same minimum of the axion inflaton potential. 

Secondly, the population of the inflationary axion valley is only known to proceed within the semi-classical regime, and within parametrically controlled approximations, via quantum tunneling, the last premise of subsection~\ref{sec:tunnel}. Entering the inflationary axion valley of large-field inflation while providing at least 60 e-folds of slow-roll inflation after tunneling thus requires tunneling from a local dS vacuum to super-Planckian inflaton vev post-tunneling. If the only local dS vacuum in the false is the one at zero inflaton VEV, then a direct Euclidean bounce with such boundary conditions is impossible, requiring the inflaton axion to first quantum diffuse uphill in the false vacuum axion valley. This leads to exponential suppression in the population of large-field inflation in string theory compared to the small-field setup contained in every large-field model via tuning generically present instanton corrections. However, the instanton correction may induce multiple local false dS vacua in the false-vacuum inflaton axion valley, while being absent in the true-vacuum valley. This is generic in the landscape, the instantons being allowed to vary in size over moduli space. Then the population of the large-field and small-field regimes can proceed from a local dS vacuum of the false-vacuum valley which is close to the 60 e-fold point of the large-field regime. Due to the absence of overshoot post-tunneling this populates the small-field and the large-field regime evenly.

The next crucial fact is the indifference of the dynamics of eternal inflation and tunneling to the vacuum energy of regions of cosmologically viable slow-roll inflation (i.e. satisfying COBE normalization) and very small-c.c.~descendant dS vacua. Both global volume-weighted and local measures combine with the high-dimensionality of the moduli space and the anthropic requirement of efficient population of an exponentially large number of such descendant vacua such, that the immediate progenitor vacua are of very high-scale to almost Planckian vacuum energy. Down-hill tunneling into the descendant vacua of parametrically small c.c.~then proceeds democratically which allows us to reduce the question of the relative prevalence of large-field and small-field inflationary regions to one of mere counting.

This counting is hard in general due to lack of calculational access. However, if we restrict the scope to first obtaining an answer for a region of the landscape with established control, counting may be feasible. As an example we gave a sketch of the discussion for the landscape of elliptically fibered 4-folds in F-theory. A large sample (several millions) of such potentially low-energy supersymmetric compactifications are fully computationally accessible in terms of hypersurfaces in toric ambient space described completely by the discrete data of the associated GLSMs. Hence, in this sector of the landscape we may be able to get an estimate of the fraction $\beta_{h^{1,1}_-\geq 1}$ of CY manifolds in the sample which have the RR-form axions required for axion monodromy large-field inflation in the first place (which we leave for future work). As such, an estimate of $\beta_{h^{1,1}_-\geq 1}<1$ would provide an upper bound on the fraction of CY's which carry axion monodromy inflation. It is conceivable that a study of random matrix models along the lines of~\cite{Marsh:2011aa} may yield in fact that small-field models are more abundant than dS minima themselves. Combined with $\beta_{h^{1,1}_-\geq 1}<1$ this would imply the absence of detectable tensor modes if a detection of low-energy supersymmetry pointed towards CY's.

Finally, if there is a way of shifting around the axion minimum as a function of the moduli \emph{without} spoiling the shift symmetry, or if there is a mechanism to protect large-field models without relying on an effective shift symmetry, the argument as it is fails.

\section*{Acknowledgments}
I am deeply indebted for many crucial and insightful discussions with R.~Bousso and E.~Silverstein. I am grateful to M.~Aganagic, A.R.~Brown, A.~Dahlen, S.~Kachru, M.~Larfors, A.~Linde, D.~L\"ust, L.~McAllister, M.~Rummel, S.~Shenker, V.~Vanchurin, and P.M.~Vaudrevange for many elucidating comments.
This work was supported by the Impuls und Vernetzungsfond of the Helmholtz Association of German Research Centres under grant HZ-NG-603, and German Science Foundation (DFG) within the 
Collaborative Research Center 676 "Particles, Strings and the Early Universe".

\newpage

\appendix

\section{Suppression of uphill tunneling}
\label{sec:App}

Here we shall shortly discuss the process of tunneling uphill from a lower-lying dS vacuum into a higher-lying one. This process is highly exponentially suppressed compared to downhill tunneling, and as a function of the vacuum energy of the final higher-lying minimum. One can see this explicitly in three different regimes of CdL tunneling. As we have the hierarchy $V_{small-field},V_{large-field}\gg V_{lowest\,dS}\simeq 0$, we can approximate the tunneling bounce by putting $V_{lowest\,dS}=0$.

\subsection{CdL tunneling in the thin-wall approximation}
\label{sec:App1}

At first we look at the case of a high potential barrier $V_T\gg V_{small-field},V_{large-field}, V_{lowest\,dS}$ which places us into the regime of the thin-wall approximation. For this situation, the Euclidean bounce action including the effects of gravity reads
\beq
S_E(\chi)=-\frac{24\pi^2}{V_+}\cdot\left[1-\frac{\big(\frac{3T^2}{4V_+}\big)^2}{\big(1+\frac{3T^2}{4V_+}\big)^2}\right]
\eeq
where $V_+=V(\chi_+)=V_{small-field}$ or $V_{large-field}$, respectively.
\beq
T=\int_{\chi_-}^{\chi_+}d\chi\sqrt{2 (V(\chi)-V_+)}
\eeq
denotes the tension of the CdL bubble wall, with $\chi_-$ denoting the position of lowest-lying dS minimum $V_-=V(\chi_-)$. The ratio $T^2/V_+$ controls the importance of the gravitational correction inside the rectangular bracket. If we approximate the potential barrier separating $\chi_\pm$ as being of height $V_T\gtrsim V_+$ and thickness $\Delta\chi$ we can write $T\sim \Delta\chi\sqrt{V_T}$. Gravity is important for $V_+\ll T^2$, or equivalently $\Delta\chi\gg\sqrt{V_+/V_T}$, resulting in
\beq
S_E^{strong\,grav.}(\chi)=-\frac{64\pi^2}{T^2}\quad.
\eeq
For sub-Planckian barrier thickness we expect that the leading terms in the scalar potential which are responsible for the two adjacent local minima at $\chi_\pm$ will also produce the barrier separating them. Therefore, if $V_+$ is not subject to specific tuning, we expect the barrier height to vary roughly together with $V_+$ as $V_+\lesssim V_T$. This leads to 
\beq
S_E^{strong\,grav.}(\chi)\sim-\frac{64\pi^2}{\Delta\chi^2\,c\,V_+}
\eeq
with some $c>0$. This is a regime where a hierarchy
\beq\label{CDLhierarchy}
S_E^{strong\,grav.}(V_+)>S_E^{strong\,grav.}(V_+')\quad,\quad V_+<V_+'
\eeq is valid. If the barrier thickness $\Delta\chi$ is sufficiently sub-Planckian, increasing $V_+$ while keeping $V_T$ fixed will eventually take us into opposite regime $\Delta\chi\ll\sqrt{V_+/V_T}$ of weak gravity where
\beq
S_E^{weak\,grav.}(\chi)=-\frac{24\pi^2}{V_+}\quad.
\eeq
Thus, we see that increasing $V_+$ eventually leads to a regime where eq.~\eqref{CDLhierarchy} is again valid. In summary, CdL tunneling in the thin-wall approximation yields a hierarchy leading to an exponential suppression of uphill tunneling scaling as $\Gamma'/\Gamma \sim \exp(-c/V_+)$ for  $V_+<V_+'$.

The exception is a situation where $V_+>0$ is tuned to be extremely small compared to the barrier height $V_T$. However, this limit is irrelevant for the discussion here, as the discrimination between large-field and small-field inflation around $\Delta\phi_{60}\simeq M_{\rm P}$ corresponds to a change of the inflationary potential energy by about 2 orders of magnitude around the GUT scale.

\subsection{CdL tunneling away from the thin-wall approximation}
\label{sec:App2}

We saw in the last section how lifting $V_+$ towards $V_T$ shuts down the gravitational correction in the thin-wall limit. However, eventually this limit will also leave the thin-wall approximation itself. There are no general explicit results for the bounce action away from the thin-wall approximation known for generic potentials. However, one may approximation any given smooth potential with two local minima by triangulating it with linear functions. Coleman tunneling in such an approximative piecewise linear potential can be solved exactly by analytical methods without using the thin-wall approximation~\cite{Duncan:1992ai}. In the case where $\Delta V_+ < \Delta V_-/4$, the bounce action can be found to be~\cite{Duncan:1992ai}
\beq
S_E(\chi)=\frac{32\pi^2}{3}\cdot\frac{1+c}{(\sqrt{1+c}-1)^4}\cdot\frac{\Delta\chi_+^4}{\Delta V_+}\quad.
\eeq
Here it is
\beq
c=\frac{\Delta V_-}{\Delta V_+}\cdot\frac{\Delta\chi_+}{\Delta\chi_-}\quad,\quad\Delta V_\pm=V_T-V_\pm\quad,\quad \Delta\chi_\pm=\pm (\chi_T-\chi_\pm)\quad.
\eeq
For constant barrier thickness parameters $\Delta\chi_\pm$ taking $V_+\to V_T$ implies $\Delta V_+\to 0$, and $c\gg 1$ which yields
\beq
S_E(\chi)\simeq \frac{32\pi^2}{3}\frac{\Delta\chi_+^3\Delta\chi_-}{V_+}\quad.
\eeq
From the discussion of the gravitational correction factor in the regime of $V_+\lesssim V_T$ of the last section we expect gravitational corrections to the flat space result just given to be small. In summary, we again find the hierarchy demanded by eq.~\eqref{CDLhierarchy} which leads to an exponential suppression of uphill tunneling scaling as $\Gamma'/\Gamma\sim \exp(-c/V_+)$ for  $V_+<V_+'$.

\subsection{Hawking-Moss tunneling - the 'no-wall' limit}
\label{sec:App3}

Finally, we can discuss tunneling in the limit of a wide and flat barrier with $V''(\chi_T)/H(V_T)^2<1$. This process is mediated by the Hawking-Moss instanton, and can be understood as uphill quantum diffusion of the scalar field $\chi$, see the end of section~\ref{sec:tunnel}. In our context of uphill tunneling from $V_-$ towards $V_+>V_-$ this gives a tunneling rate (see eq.~\eqref{HMProb})
\beq
\Gamma_{HM}\sim e^{-\left(\frac{1}{V_-}-\frac{1}{V_T}\right)}\quad.
\eeq
The ratio of tunneling rates for tunneling uphill from $V_-$ into two different higher-lying vacua with vacuum energies $V_+\lesssim V_T\ll V_+'\lesssim V_T'$ thus comes out to be
\beq
\frac{\Gamma_{HM}'}{\Gamma_{HM}}\sim e^{\frac{1}{V_T'}-\frac{1}{V_T}}\sim e^{-\frac{1}{V_T}}\quad.
\eeq
Again, we find the hierarchy of eq.~\eqref{CDLhierarchy}, and therefore tunneling uphill is severely punished for increases in the potential energy $V_+\lesssim V_T$ of the tunneling destination.

\section{Progenitor dS vacua - global vs. local measures of eternal inflation}
\label{sec:AppB}

There are quite a number of measures of eternal inflation which have been proposed to this date (for a by no means complete list of recent works see e.g.~\cite{Bousso:2006ev,Freivogel:2006xu,Linde:2006nw,Clifton:2007en,Linde:2007nm,Clifton:2007bn,Creminelli:2008es,Winitzki:2008yb,Bousso:2008as,DeSimone:2008if,Garriga:2008ks,Linde:2008xf,Freivogel:2009it,Freivogel:2009rf,Bousso:2009mw,Garriga:2009hy,Linde:2010xz,Bousso:2010yn,Vilenkin:2011yx,Guth:2011ie,Dubovsky:2011uy,Harlow:2011az}), even if we only count those which do not immediately run into paradoxa such as the youngness, or the Boltzmann brain problems. However, for the purpose of our discussion their most important property is that they separate into two classes with respect to the one relevant aspect here -- the typical vacuum energy of the progenitor dS vacua. The discussion of this aspect follows in particular~\cite{Linde:2006nw}.

\begin{itemize}

\item $(G)$ One class (the so-called global measures) rewards different inflationary vacua or regions of the landscape proportional to the volume of 3-space generated during the eternal phase. Correspondingly, in volume-weighted global measures all the vacua and slow-roll inflationary regions of the landscape are seeded ultimately by the highest-lying meta-stable dS vacuum of the landscape.

Let us illustrate this at the example of a simple toy landscape with 3 vacua $V_{\rm s}<0<V_1< V_2$ which is depicted in Figure~\ref{landscape}. This example features the general structure of the string landscape -- neighbouring vacua tend to have large differences in vacuum energy, the width of the potential barriers are $\lesssim M_P$, and there are AdS vacua.
\begin{figure}[ht]
\begin{center}
\includegraphics[width=0.9\textwidth]{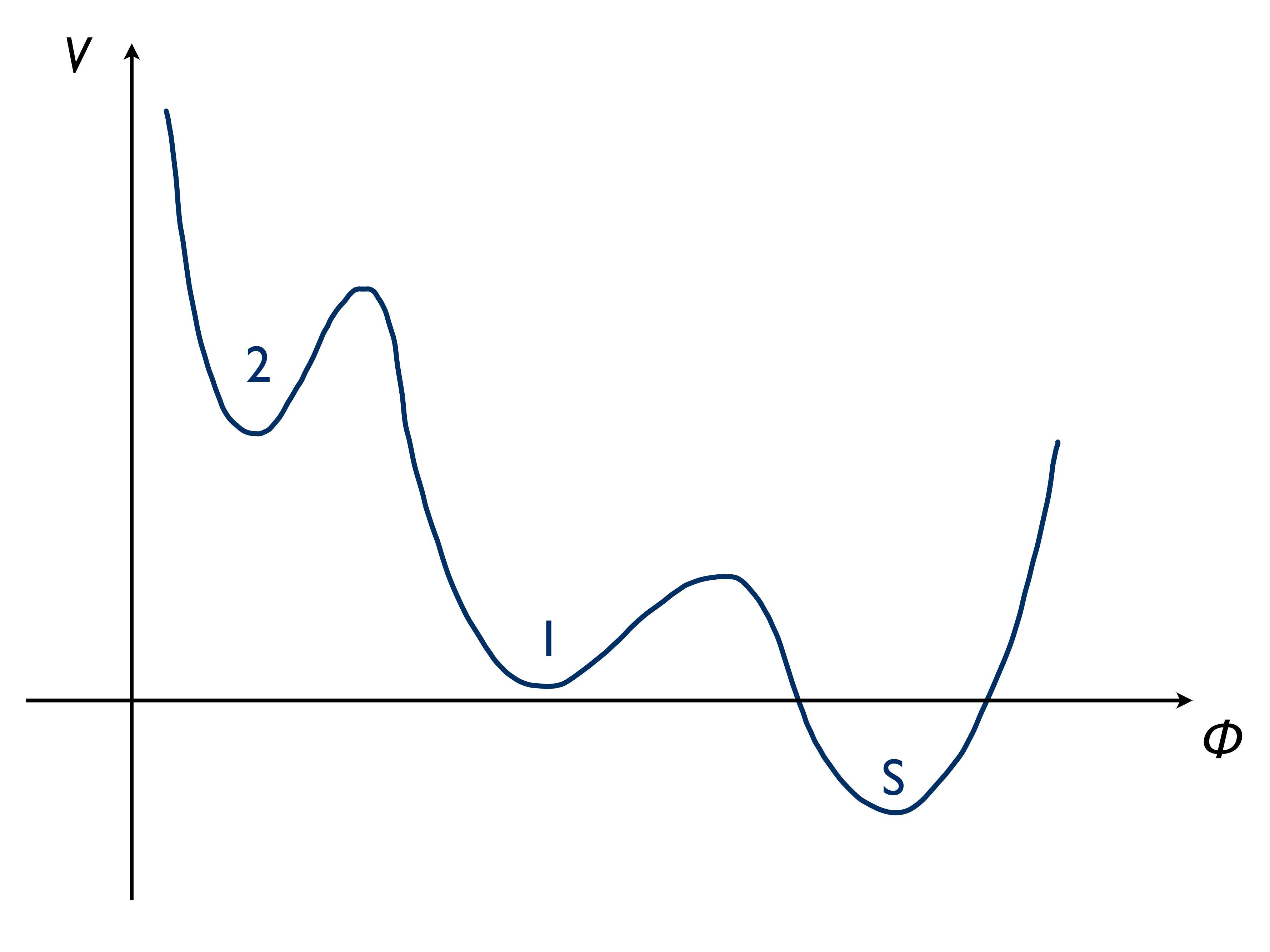}
\end{center}
\refstepcounter{figure}\label{landscape}
\vspace*{-.2cm} {\bf Figure~\ref{landscape}:} A simple toy landscape with 3 vacua $V_{\rm s}<0<V_1< V_2$. $V_{\rm s}$ is an AdS vacuum and acts as a so-called 'sink', i.e. a vacuum where eternal inflation ends in a crunch.
\end{figure}
Therefore, we choose the vacuum $S$ to be an AdS vacuum of negative cosmological constant $V_s<0$ . Tunneling from vacua with positive vacuum energy, such as the vacua $1$ and $2$ into vacuum $S$ will create AdS bubbles within which space-time ends in a big brunch. Therefore, the AdS vacuum acts as a sink, destroying probability current flowing from the eternal inflating vacua $1$ and $2$.

The vacuum population dynamics of this system is governed by differential rate equations. They determine the rate of change of probability $\dot P_i$ of realizing vacuum $i$ by the probability currents $J_{ij}$ which feed or drain vacuum $i$~\cite{Linde:2006nw,Clifton:2007en}
\bea\label{rateeqn1g}
\dot P_1 &=& -J_{1s}-J_{12}+J_{21}+\,J_{1,vol}\nonumber\\
&&\\
\dot P_2 &=& -J_{2s}-J_{21}+J_{12}+\,J_{2,vol}\quad.\nonumber
\eea
The probability currents are given as $J_{ij}=P_i \Gamma_{ij}$, and $J_{i,vol}=P_i\cdot 3H_i$. Here, $\Gamma_{ij}$ denotes the decay rate for forming bubbles of vacuum $j$ in a sea of vacuum $i$. Note, that in a global measure the volume growth $\sim e^{3 H_i t}$ of each vacuum $i$ is weighted for by adding $J_{i,vol}$. The dS-dS vacuum decay rates are given from CdL tunneling as
\beq
\Gamma_{21}=e^{-S(\phi)+S_2}\quad,\quad \Gamma_{12}=e^{-S(\phi)+S_1}\quad,\quad S_i\equiv S(\phi_i)=-\,\frac{24\pi^2}{V_i}
\eeq
while we denote the decay of vacuum 1 into the AdS vacuum $S$ by $\Gamma_{1s}=e^{-C_1}$. From now on, we will set $\Gamma_{2s}=0$ for simplicity. Then the vacuum dynamics reads
\bea\label{rateeqn2g}
\dot P_1 &=& -P_1 (\Gamma_{1s}+\Gamma_{12})+P_2 \Gamma_{21}+3H_1\,P_1\nonumber\\
&&\\
\dot P_2 &=& -P_2 \Gamma_{21}+P_1\Gamma_{12}+3H_2\,P_2\quad.\nonumber
\eea

We will assume $\Gamma_{21}\gg \Gamma_{12}$ as usually $V_2>V_1$, i.e. up-hill tunneling is highly suppressed. Furthermore, in most cases we have overwhelmingly $H_i\gg \Gamma_{ij}$.

With these inputs, eq.s~\eqref{rateeqn2g} has a solution~\cite{Linde:2006nw}
\beq
\frac{P_2}{P_1}=\frac{3 (H_2-H_1)}{\Gamma_{21}}\gg 1\quad,\quad P_1\sim e^{3 H_2 t}\quad,\quad P_2\sim \frac{3 (H_2-H_1)}{\Gamma_{21}}\,e^{3 H_2 t}\quad.
\eeq
All vacua grow with the volume growth of the highest-lying meta-stable dS vacuum whose population dominates everything else. This does not depend on the decay rate $\Gamma_{1s}$ into the AdS sink, as long as $H_2> \Gamma_{1s}$.

\item $(L)$ Conversely, the other class (the local measures) discards rewarding the 3-space volume generated. They just account for the bare-bones anthropically required 60-odd e-folds of slow-roll volume growth. Local measures seed all vacua from the longest-lived progenitor.

In this case, the vacuum dynamics is governed by
\bea\label{rateeqn1L}
\dot P_1 &=& -P_1 (\Gamma_{1s}+\Gamma_{12})+P_2 \Gamma_{21}\nonumber\\
&&\\
\dot P_2 &=& -P_2 \Gamma_{21}+P_1\Gamma_{12}\quad.\nonumber
\eea
Note that the volume growth rate contributions are absent by definition of the local nature of this class of measures. There is one variation to this class of measures, in that there is a local-global duality which links the local causal patch measures with the global 'scale-factor' measure. The scale-factor measure adds back volume growth terms $J_{i,vol}=3 P_i$
\bea\label{rateeqn1LsfM}
\dot P_1 &=& -P_1 (\Gamma_{1s}+\Gamma_{12})+P_2 \Gamma_{21}+3P_1\nonumber\\
&&\\
\dot P_2 &=& -P_2 \Gamma_{21}+P_1\Gamma_{12}+3P_2\quad.\nonumber
\eea

However, these lead to universal volume growth $\sim e^{3t}$ of all dS vacua. This implies, that the overall volume growth can factored out unambiguously, so that ratio of vacuum population probabilities behave exactly as in local causal patch measures (this is a manifestation of the 'global-local duality' between causal patch measures and the scale-factor measure~\cite{Bousso:2009mw}).

The asymptotic behaviour of the ratio $P_2/P_1$ does now have two distinct regimes, depending on whether $\Gamma_{1s}\ll \Gamma_{21}$ (a 'narrow' sink) or the opposite  $\Gamma_{1s}\gg \Gamma_{21}$ (a 'wide' sink) is realized~\cite{Linde:2006nw,Clifton:2007en}. For a narrow sink we find
\beq
\frac{P_2}{P_1}=\frac{\Gamma_{12}}{\Gamma_{21}}=e^{S_1-S_2}=e^{\frac{24\pi^2}{V_2}-\frac{24\pi^2}{V_1}}\ll1\quad.
\eeq
The opposite case of a wide sink yields
\beq
\frac{P_2}{P_1}=\frac{\Gamma_{1s}}{\Gamma_{21}}=e^{-S_2+S(\phi)-C_1}\approx e^{-S_2+S(\phi)}\approx e^{\frac{24\pi^2}{V_2}}\gg 1\quad.
\eeq
Both cases share a common property -- for a narrow sink, the vacuum populations are dominated by vacuum 1, while for a wide sink vacuum 2 dominates -- in each cases it is the \emph{longest-lived} dS vacuum which dominates the landscape in the stationary limit, and in turn then feeds everyone else~\cite{Linde:2006nw,Clifton:2007en}.
\end{itemize}

\bibliographystyle{JHEP.bst}
\bibliography{tensors}
\end{document}